\documentclass[useAMS,usenatbib,fleqn]{mn2e}

\usepackage{graphicx, amssymb}
\usepackage[fleqn]{amsmath}
\usepackage{epsfig}
\usepackage{color}
\usepackage{times}

\newcommand{\beq}{\begin{equation}}
\newcommand{\eeq}{\end{equation}}
\newcommand{\bea}{\begin{eqnarray}}
\newcommand{\eea}{\end{eqnarray}}

\title[Transport and Heterogeneity in Early Star Forming Systems]{Metal Transport and
  Chemical Heterogeneity in Early Star Forming Systems}

\author[J. Ritter et al.]
{Jeremy S.~Ritter, Alan~Sluder, Chalence~Safranek-Shrader, Milo\v
  s~Milosavljevi\'c \newauthor  and Volker~Bromm\\
Department of Astronomy, The University of Texas at Austin, Austin, TX 78712, USA}
 
\newcommand{\apj}{ApJ}

\newcommand{\mnras}{MNRAS}
\newcommand{\araa}{ARA\&A}
\newcommand{\apjs}{ApJS}
\newcommand{\nat}{Nature}
\newcommand{\apjl}{ApJ}
\newcommand{\aap}{A\&A}

\newcommand{\pasp}{PASP}

\begin{document}

\label{firstpage}

\maketitle
\topmargin-1cm

\begin{abstract}

To constrain the properties of the first stars with the chemical abundance patterns
observed in metal-poor stars, one must
identify any non-trivial effects that the hydrodynamics of metal
dispersal can imprint on the abundances. We 
use realistic cosmological hydrodynamic simulations to quantify the distribution 
of metals resulting from one Population III supernova and from a small
number of such supernovae exploding in close succession.  Overall, supernova ejecta
are highly inhomogeneously dispersed throughout the simulations.  When the
supernova bubbles collapse, quasi-virialized
metal-enriched clouds, fed by fallback from the bubbles and
 by streaming of metal-free gas from the cosmic web, 
grow in the centers of the dark matter halos.  Partial turbulent
homogenization on scales resolved in the simulation is observed only in the densest
clouds where the vortical time scales are short enough to ensure true
homogenization on subgrid scales. However, the abundances in
the clouds differ from the gross yields of the supernovae.  Continuing the simulations until
the cloud have gone into gravitational collapse, we predict
that the abundances in second-generation stars will be
deficient in the innermost mass shells of the supernova (if only one
has exploded) or in the ejecta of the latest supernovae (when multiple
have exploded).  This indicates that 
hydrodynamics gives rise to biases complicating the identification of 
 nucleosynthetic sources in the chemical abundance spaces of the surviving
stars.

\end{abstract}


\section{Introduction}
\label{sec:intro}

The chemical abundance patterns of metal-poor and ancient
stellar populations and 
intergalactic absorption systems provide information about the
earliest stages of
chemical enrichment \citep[for reviews, see][]{Beers:05,Frebel:13a,Karlsson:13,Frebel:15}.  Because the first
stellar systems were likely enriched by only a few
discrete sources, one must interpret the inherent complexity of
enrichment in relating the abundance patterns measured in metal-poor stars
to theoretical models \citep{Audouze:95,Shigeyama:98}. Cosmic star formation, the
driver of enrichment, is already stochastic thanks to the randomness of cosmic
primordial density fluctuations. Significant additional complexity
enters chemical
enrichment through the
turbulent hydrodynamics of the intergalactic, circumgalactic, and interstellar medium, the stochasticity
of the star formation process, and 
the mechanics of the dispersal of
nucleosynthetic products from their sources---supernovae and stellar
mass loss.  The effects of complexity are partially, though perhaps not completely erased through
the stirring and mixing in mature star forming systems like the Milky Way's disk. 
But in ancient stellar populations, we expect vestiges of the complexity to
persist, requiring us to model it as
much as possible from first principles.  

The recent years have seen a class of high-resolution simulations of small numbers of primordial star forming systems
\citep[e.g.,][]{Wise:08,Greif:10,Wise:12,Wise:14,Ritter:12,Jeon:14} as well as of moderate-resolution simulations (somewhat farther removed from the first-principles philosophy) that allow the authors to track large samples of such systems \citep[e.g.,][]{Ricotti:08,Bovill:09,Tassis:12,Muratov:13,OShea:15}.  Without exception, these works study the local enrichment amplitude as quantified by the metallicity but do not attempt to investigate potential complex  effects that could influence differential chemical abundance patterns.

A critical source of uncertainty entering
chemical evolution models is the chemical yields of
supernovae. 
The yields could vary drastically with the progenitor mass, rotation
rate, the presence of a binary companion, and initial metal content \citep[e.g.,][]{Heger:02,Heger:10}.  
In the case of core-collapse supernovae, 
first-principles theoretical modeling is still not
sufficiently predictive to permit direct synthesis of the observed
stellar chemical abundance patterns. What is worse, the yields may not be deterministic and 
could be sensitive to
instabilities taking place immediately prior to and in the course of
the explosion \citep[e.g.,][]{Arnett:11,Ellinger:12,Wongwathanarat:13,Smith:14,Couch:15}.
It has therefore been a longstanding hope that the process of
nucleosynthesis will be reverse engineered from the abundance
patterns \citep[e.g.,][]{Talbot:73,Nomoto:13}.  Principal component
analysis (PCA) can be applied in the stellar chemical abundance space
to discern the contributing classes of nucleosynthetic sources, be they
individual stars or star clusters
\citep[e.g.,][]{Ting:12}, if one posits that abundance patterns of the
sources define monolithic basis vectors of the resulting stellar chemical
abundance space.   This is becoming a particularly promising direction with the arrival of large spectroscopic surveys such as HERMES-GALAH \citep{Zucker:12,DeSilva:15} and Gaia-ESO \citep{Gilmore:12} surveys.   However, the applicability of PCA becomes less clear if the 
hydrodynamics of metal dispersal skews the patterns,
introducing biases deviating the patterns from linear superpositions of the patterns of well-defined nucleosynthetic source classes.  Properly characterizing
such biases would place the recovery of the properties of
nucleosynthetic sources from stellar abundance data on much firmer footing.

The most primitive and metal-poor stellar populations known are the ultra-faint dwarf
spheroidal satellites of the Milky Way \citep[UFDs;][]{Brown:12,Frebel:12,Vargas:13,Frebel:14}, the Milky
Way's stellar halo which is thought to have formed (at least in part) through the
disruption of dwarf satellite galaxies (\citealt{Kirby:08}, \citealt{Norris:10a}, \citealt{Lai:11}, but see, e.g., \citealt{Lee:13}) 
and possibly an old, metal-poor sub-population within the Milky Way bulge.\footnote{We can define a stellar population as ``primitive'' if the number of times the nucleosynthetic output of individual supernovae and AGB stars has been recycled is small.}
The prominent chemical heterogeneity in UFDs is often invoked as a potential indicator of poor mixing of supernova ejecta or an enrichment by a small number of contributing supernovae \citep[e.g.,][]{Norris:10c,Simon:15}.
The primitive populations could be
particularly sensitive to hydrodynamical biases, given the low masses and
gravitational potential well
depths of the progenitor star-forming systems and the low numbers of
contributing supernovae and AGB stars.

Here we make an initial attempt
to identify these biases with ultra-high-resolution
cosmological hydrodynamic simulations. We track the degree of 
chemical heterogeneity in the enrichment by a single supernova, and by a
cluster of seven consecutive supernovae.  The supernovae explode in a previously
metal-free cosmic minihalo, a plausible UFD progenitor \citep{Ricotti:08,Salvadori:09}.  The two
simulations are initialized from Gaussian cosmological fluctuations,
ensuring that the cosmic environment being enriched is realistic.  

The finite resolution of the simulations implies that we are able to directly
place only a lower limit on the degree of \emph{coarse-grained} chemical
heterogeneity. Perfect chemical homogeneity, down to molecular scales, is expected only if the gas
has been stirred by such processes
as gravitational infall, the mechanical feedback from star formation,
and thermal instability, to the extent that would facilitate microscopic
diffusion across chemically heterogeneous sheets produced by
turbulent folding.  Efficient
stirring generally requires that the vortical time $\sim |{\bf
  \nabla}\times{\bf v}|^{-1}$ be much shorter than the lifetime of a cloud before it collapses to form new stars \citep[see, e.g.,][and references therein]{Pan:10}.
This condition can be fulfilled in specific sites, but
certainly not in general.\footnote{Microscopic chemical mixing is a
sufficient, but not a necessary condition for chemical homogeneity,
since any residual heterogeneity on sub-stellar mass scales is erased
in the stars themselves.}  

We study the transport of supernova
ejecta resolved by the mass coordinate inside the progenitor star (in
the simulation with one supernova) and by the temporal order of the explosion (in
the simulation with multiple supernovae).  We follow the transport until gas clouds have assembled in which 
second-generation stars can be expected to form.  This allows us to determine if these
nucleosynthetic sources contribute monolithically, defining invariant basis
vectors of the chemical abundance space, or if, perhaps, the
hydrodynamics of supernova remnant evolution favors the reprocessing
into new stars of only a biased subsample of the gross nucleosynthetic
yields, skewing toward the reprocessing of some
elements but not others.

\begin{figure*}
\begin{center}
\includegraphics[width=0.33\textwidth]{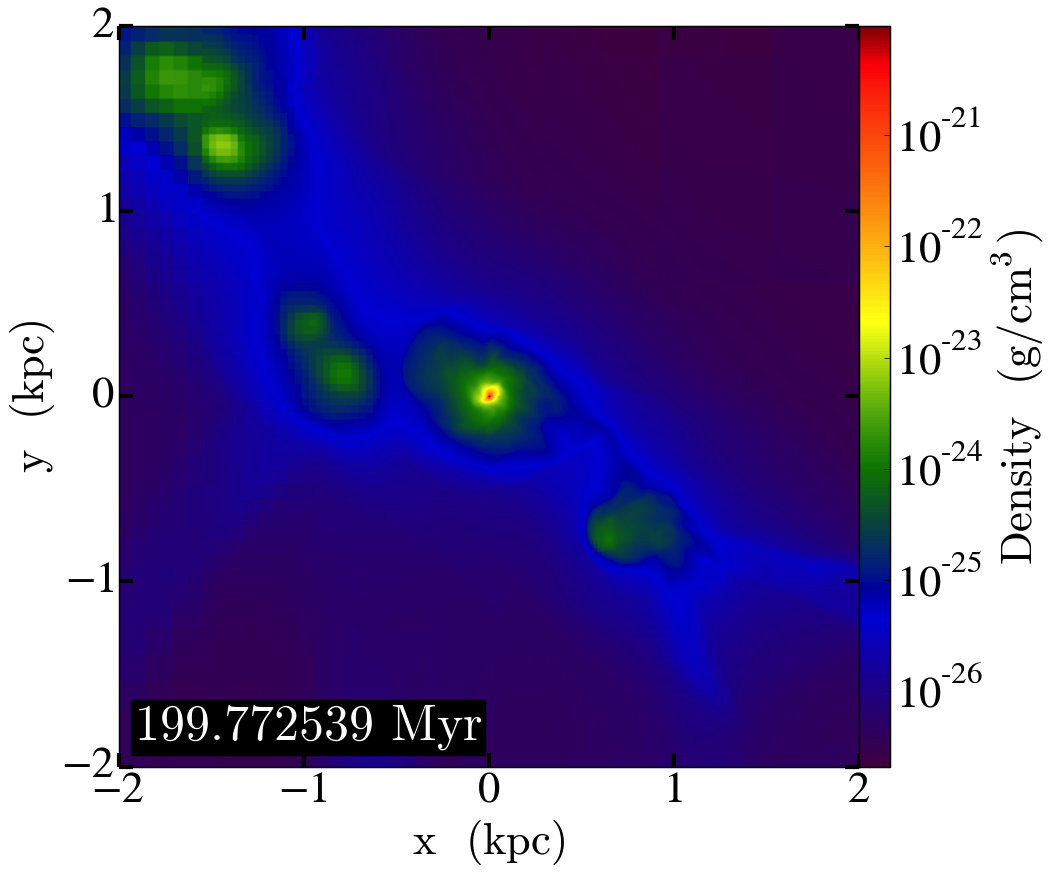}%\hspace{0.5cm}
\includegraphics[width=0.33\textwidth]{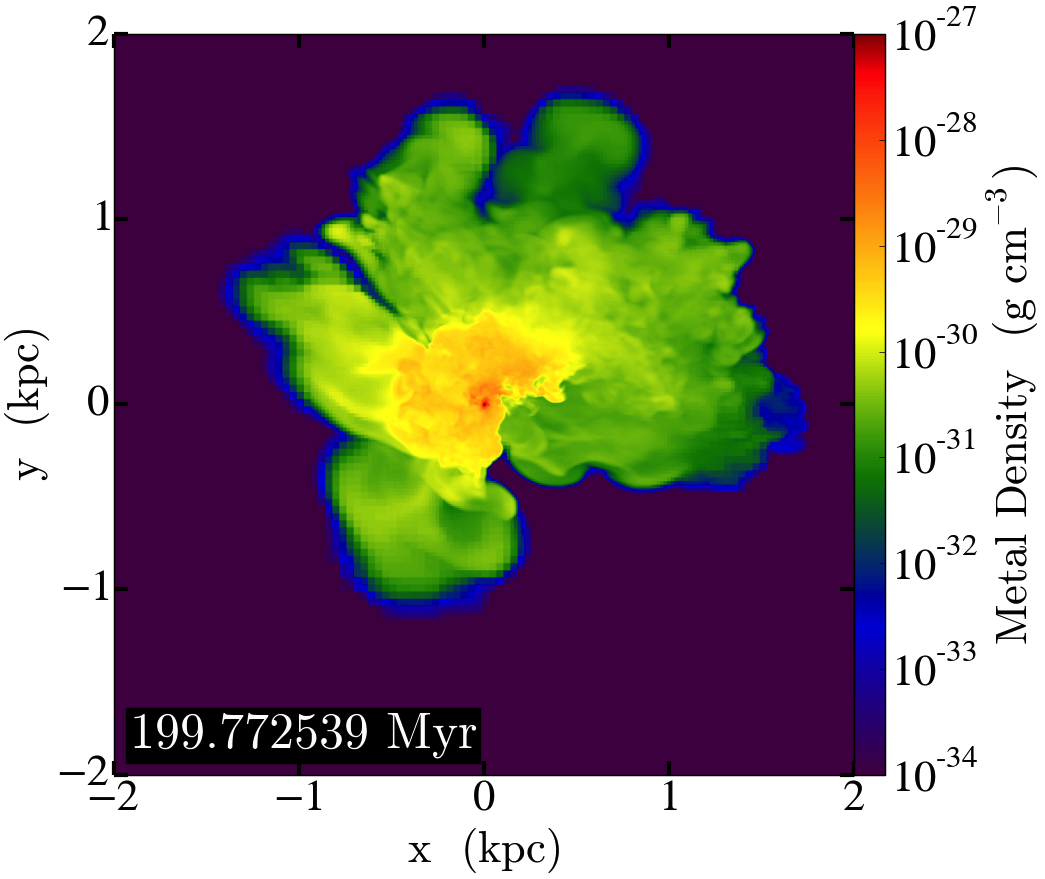}
\includegraphics[width=0.33\textwidth]{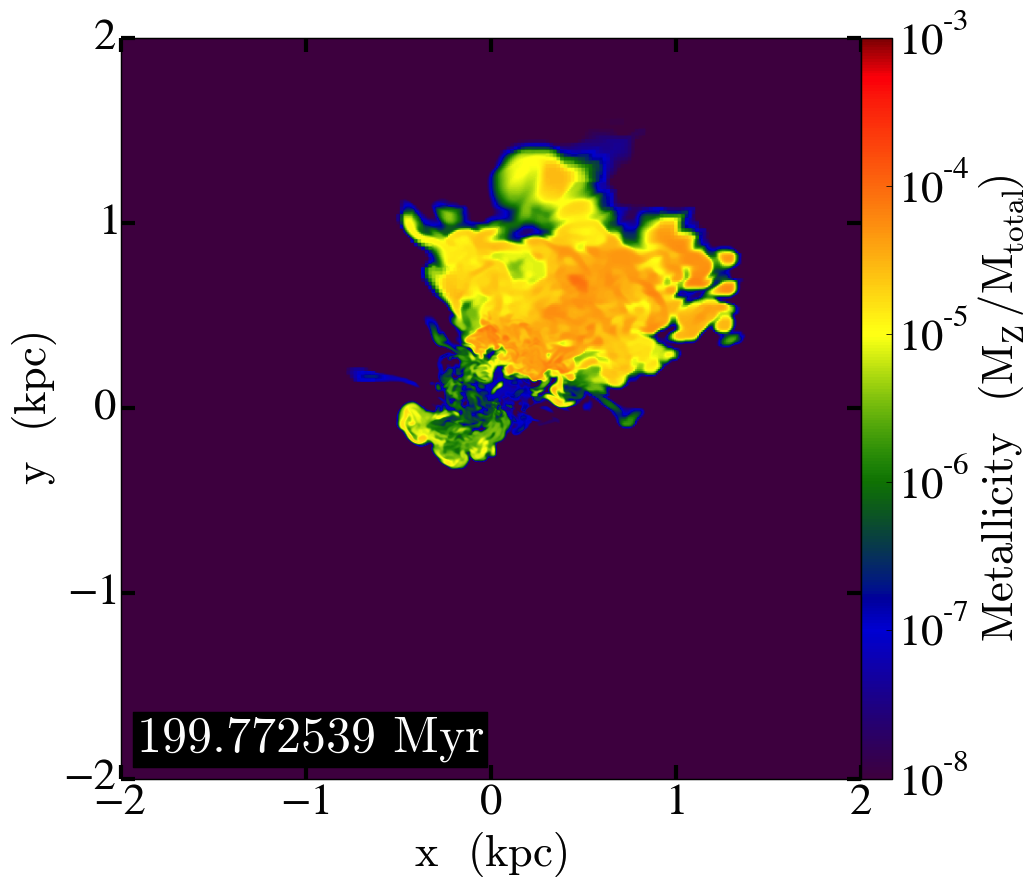}
\end{center}
\caption{Simulation \textsc{7sn} $200\,\textrm{Myr}$ after the first
  supernova explosion: density-weighted density projection
  ($\int\rho^2dz/\int\rho dz$, left),
metal-density-weighted metal density projection
($\int\rho_{Z}^2dz/\int\rho_Z dz$,  middle), and metallicity slice
through the cell with maximum density (right). The metal density is derived
from the advected passive scalar tracing supernova ejecta. The cosmic
web filament extends diagonally from top left.  Note the relatively
low metallicity near the center, resulting from a dilution of supernova
ejecta with metal-free gas streaming down the cosmic web filament.\label{fig:7sn_projections_slices}}
\end{figure*}

\section{Numerical Methodology}
\label{sec:method}

The simulations were initialized from the same realization of
cosmological initial conditions as in \citet{SafranekShrader:12} and \citet{Ritter:12}.  
%The methods were similar to and improvements of those in
%\citet{Ritter:12} and will be described in detail in follow-up papers.
We simulated the gravitational collapse of collisionless dark
matter and baryonic fluid in a box of size $1\,\textrm{Mpc}$ 
(comoving) starting at redshift $z=146$. The initial density and
velocity perturbations were generated with the multiscale cosmological 
initial conditions package \textsc{grafic2} \citep{Bertschinger:01} with 
Wilkinson Microwave Anisotropy Probe 7-year cosmological parameters \citep{Komatsu:11}. We utilized
two levels of nested refinement to achieve an effective resolution of
$512^3$, corresponding to $230\,M_\odot$ per dark matter particle, in
a patch encompassing the density maximum on mass scales $10^8\,M_\odot$. Time
integration was carried out with the adaptive mesh refinement (AMR)
code \textsc{flash} \citep{Fryxell:00} as described in \citet{SafranekShrader:12}. 
In what follows, we
use physical units as opposed to comoving units.  Metallicities
are quoted in absolute metal mass fractions unless specified otherwise.

Gas cooled by ${\rm H}_2$ rovibrational emission collapsed to form a $\approx 10^6\,M_\odot$
minihalo at redshift $z \sim 19.6$. We inserted a single collisionless particle
at the location of the gas density maximum within the minihalo
to represent a Population III star (simulation \textsc{1sn}) or a small
cluster of such stars (simulation \textsc{7sn}).
%When gas in a $\approx 10^6\,M_\odot$ minihalo, cooled by 
%${\rm H}_2$ rovibrational emission, has collapsed to density
%$\gtrsim 10^3\,{\rm cm}^{-3}$, a single collisionless particle representing a
%Population III star (simulation \textsc{1sn}) or a small cluster of
%such stars (simulation \textsc{7sn}) was
%inserted.  
The equivalent gas mass was removed by reducing
the nearby gas density to a constant maximum level.  
In view of the recent realization that protostellar disk fragmentation \citep{Stacy:10,Clark:11a,Clark:11b,Greif:11,Greif:12}
and evaporation by protostellar radiation \citep{Hosokawa:11,Stacy:12}
can limit the masses of Population III stars in the few tens of solar
masses, we
picked their masses to be in the range $20-80\,M_\odot$ with
typical values $\sim40\,M_\odot$ and assumed for simplicity that they all exploded with energies
$10^{51}\,\textrm{erg}$.  Preceding the first explosion, we let 
the collisionless particle emit ionizing radiation
for $3\,\textrm{Myr}$ with an ionizing luminosity $Q_{\rm ion}$ and create an H\,II region. 
Hydrodynamic expansion of the ionized gas reduced the central gas density to $n\sim
0.3\,\textrm{cm}^{-3}$.  After $3\,\textrm{Myr}$, we either inserted a single supernova
remnant (\textsc{1sn}; $Q_{\rm ion}=6\times10^{49}\,\textrm{photons}\,\textrm{s}^{-1}$) or initiated a sequence of 7
consecutive supernovae (\textsc{7sn}; $Q_{\rm ion}=2.2\times10^{50}\,\textrm{photons}\,\textrm{s}^{-1}$), all centered on the
location of the collisionless particle.

In \textsc{1sn} we excised from
the cosmological simulation a
$1\,\textrm{kpc}$ region centered on the collisionless particle,
replacing the dark matter particles with a simple, parametric, spherically
symmetric, time-dependent 
analytical dark matter density profile. The analytical profile was centered in the initial, instantaneous local rest frame of the host dark matter halo.
The baryonic density and velocity field from the
cosmological box was mapped directly onto the excised region, 
allowing us to continue the simulation in the interior of the excised
region at high spatial resolution. The supernova ejecta mass was set to
$40\,M_\odot$. 
In \textsc{7sn}, the seven supernova delay times $3.1-7.7\,\textrm{Myr}$
(measured after
collisionless particle insertion) were selected to represent the
lifetimes of stars with masses decreasing from $80$ to $20\,M_\odot$, following  
a stellar IMF with a flat $dN/d\ln M$.  Each of the supernovae was inserted in the free expansion
phase, with $E_{\rm SN}=10^{51}\,\textrm{erg}$ in kinetic energy and an
initial radius smaller than one-tenth of the radius containing
gas mass equal to that of the ejecta.  
The ejecta masses were in the range
$25$--$18.5\,M_\odot$, decreasing
with the order of the explosion.  

We terminated each simulation when sufficient gas returned to
the halo center to form a self-gravitating cloud with density
$>10^3\,\textrm{cm}^{-3}$, which happened $56$ and $198\,\textrm{Myr}$
after collisionless particle insertion in simulations \textsc{1sn} and
\textsc{7sn}, respectively.

%  and $>10^4\,\textrm{cm}^{-3}$

\subsection{Thermodynamic Evolution}
\label{sec:thermodynamic}

Preceding the insertion of the collisionless star
particle we integrated the full set of coupled rate equations of the nonequilibrium chemical network for the
primordial chemical species H, He, and D, their common ions, 
and the molecules H$_2$ and HD \citep{SafranekShrader:12}.
After insertion of the star particle we turned on an ionizing point
source. We mapped the gas
density onto a source-centered spherical grid partitioned using the \textsc{healpix} \citep{Gorski:05}
algorithm in the angular coordinate (3000 pixels) and logarithmically in the radial coordinate (100 radial bins). 
The flux in any spherical cell was diminished by the cumulative extinction in the cells at smaller radii. 
We then mapped the flux from the spherical grid back onto the Cartesian simulation grid.
On the basis of the mapped flux, we determined if a cell was expected to be ionized.  
We assumed photoionization equilibrium in ionized cells 
and interpolated the local thermodynamic state of the gas from data
tabulated with the code \textsc{cloudy} \citep{Ferland:13} as a function of the 
ionization parameter proportional to the 
local hydrogen-ionizing flux divided by the total hydrogen 
number density.\footnote{While not photon-conserving, this method is sufficient to simulate the anisotropic H\,II region expansion in the relatively dense gas of our minihalo.}

When the first supernova remnant was inserted, we switched off the photoionizing
source and began computing the cooling rates assuming that the species' abundances were in 
collisional ionization equilibrium.
Here, the cooling rate and the ionization state were again interpolated from tables pre-computed with 
\textsc{cloudy}, now as a function of density, temperature, and metallicity. 
For the interpolation we used the metallicity as defined by the passive mass scalar metallicity on the Cartesian grid.  
Artificial diffusion (or, better, ``numerical teleportation'') of the passive scalar metallicity is a
problem hitherto unsolved in Eulerian codes \citep[see, e.g., Footnote 6 in][]{Ritter:12}. It produces pseudo-exponential low-metallicity tails ahead of advancing metallicity fronts.  In the hot, shocked shell preceding the supernova ejecta, the teleportation tail is present, but the metallicity in the tail is too low to contribute to gas cooling.  In gas with temperature $\gg 10^4\,\mathrm{K}$, metal cooling begins to dominate only at metallicities $\gtrsim 0.1\,Z_\odot$, higher than in the artificial tail.  On the other hand, at low temperatures $\ll 10^4\,\mathrm{K}$, metallicities even as low as $\lesssim 10^{-3}\,Z_\odot$ (and $\sim 10^{-6}\,Z_\odot$ in the presence of dust) can define the cooling rate.

Similar to the numerical teleportation of chemical abundance scalar tracers, there is the potential of insidious mass teleportation at the contact discontinuity separating the hot, shock-heated gas of the supernova bubble from the much colder gas inside the enveloping dense cooling shell. Our calculations do not include the conduction of heat from the hot to the cold side of the interface.  The net effect of thermal conduction should either be a cooling flow condensing the hot into the cold phase, or a reverse, evaporation flow, depending on the structure of the interface.  For the specific temperature gradients and densities in our simulations, one would expect the heat conduction to produce an evaporation rather than a cooling flow \citep[e.g.,][]{Cowie:77,Draine:11}.  The simulations exhibit a pressure dip at the interface, implying a narrow cooling layer.  This arises from the artificial numerical smearing of the contact discontinuity.  Dense gas from the cold side smears over onto the hot side thus artificially lowering the cooling time in a layer one cell thick.  The cell-thick pressure dip is an under-resolved hydrodynamic feature.  The numerical artifacts that it may produce should be sensitive to the particulars of the numerical scheme such as the choice of slope limiter and approximate Riemann solver.  We will perform a systematic study of numerical behavior at contact discontinuities in radiating flows in a separate, technical paper.

Since accurate tracking of molecular processes requires 
integration of stiff and computationally expensive chemical rate equations, to accelerate the computation, we did not track the H$_2$ abundance and did not include molecular cooling after the first supernova remnant insertion.  This is a potential oversimplification as H$_2$ should form as the photo- and supernova-ionized gas recombines. Absent a high dust abundance that would drive molecule formation on grains (we do not track dust but its abundance is limited from above by the metallicity that remains low), H$_2$ abundance in the recombined gas should be 
$n_{{\rm H}_2} / n_{\rm H} \sim 10^{-3} - 10^{-2}$ \citep[e.g.,][]{MacLow:86,Shapiro:87,Kang:92}.
Molecule abundance reaches 
maximum within about a million years after gas cools below 
$10^4\,\textrm{K}$.  In our simulation, this gas has metallicity $\ll 10^{-2}\, Z_\odot$ and cooling by H$_2$ should dominate
\citep{Glover:14}, at least in a relatively narrow density range $n_{\rm H}\sim 10-100\,\mathrm{cm}^{-3}$ \citep[see, e.g., Figure 6 in][]{SafranekShrader:14a}.  Again barring a full atomic-to-molecular transition that would require abundant dust, at still higher densities $\gtrsim 100\,\mathrm{cm}^{-3}$, metal line cooling dominates even before molecular transitions thermalize at $\sim10^4\,\mathrm{cm}^{-3}$.  

The artifact of our neglecting molecular cooling at densities $\sim 10-100\,\mathrm{cm}^{-3}$ is that the supernova-enriched, recollapsed gas will linger longer under quasi-hydrostatic conditions; a longer interval will pass before gravitational potential well becomes deep enough to compress the quasi-isothermal gas to the threshold of runaway gravitational collapse.
The results that follow should be interpreted as placing an upper limit on the cooling and recollapse time in a
supernova-enriched dark matter minihalo following a single supernova or a cluster of
supernovae.  The limit is strictly valid only in the fine-tuned regime in which molecule formation is completely suppressed by a dissociating UV background.  The true recollapse time could be factor of a few shorter than in our simulations if the molecular abundance is significant.  We will specifically address molecular cooling in our next work on this topic.

\subsection{Adaptive Mesh Refinement Control}

Throughout the
simulation, the AMR resolution was dynamically and adaptively adjusted to resolve the structures of interest. The refinement level $\ell$ of AMR
blocks relates the size of the computational box $L$ to size of an
individual AMR grid cell by $\Delta x = 2^{-\ell+1-3} L$, where
the factor of $2^{-3}$ arises from the subdivision of AMR blocks into $8$ cells
along each axis ($8^3 = 512$ cells per block). During the initial
gravitational collapse of dark matter and gas leading to the formation of a
minihalo we refined based on the gas density $\rho$. Specifically we raised the refinement to level $\ell$ to satisfy the condition $\rho < 3\bar \rho \, 2^{3(1+\phi)(\ell-\ell_{\rm base})}$, 
where $\ell_{\rm base} = 5$ is the initial refinement level at the start of
the simulation, $\bar\rho$ is the mean density in the box, and $\phi = -0.3$ \citep[see, e.g.,][and references therein]{SafranekShrader:12}.
After the insertion of a star particle, which takes place in the most massive dark matter
minihalo in the box, we discontinued enforcing density-based refinement in other halos.  At the evolving location of the star particle we maintained the
maximum refinement level attained when refining based on gas density even after the H\,II region broke out and gas density dropped. 

In order the resolve the free expansion phase of a supernova remnant, it
is necessary to resolve scales much smaller than the radius at which the
expanding blast wave sweeps up a gas mass similar to the ejecta mass. 
The grid resolution at each supernova insertion was forced to be $\leq 0.03\,\textrm{pc}$ and somewhat
coarser for the later supernovae in \textsc{7sn} that explode in
the bubble containing gas shocked by early supernovae.  We additionally
refined the grid up to the maximum resolution anywhere in the box using the standard
second-derivative refinement criterion in \textsc{flash} tuned to 
aggressively refine ahead of compositional discontinuities where metallicity jumps. This ensured that
the forward shock wave, contact discontinuity, and reverse shock were
maintained at the highest available grid refinement level.
As each remnant expanded, we degraded the
resolution while ensuring that the diameter of the remnant (or superbubble) was
resolved by at least 256 cells.  In the late stages of the
simulations focused on the recollapse in the halo, we degraded the resolution in regions
causally disconnected from the halo center, outside the virial radius of the halo.

In all supernovae we assumed that $10\%$ of the ejecta mass was in
metals, with $\alpha$-enhanced solar abundances
($[\alpha/\textrm{Fe}]=0.5$).  In the thermodynamic
calculations (but not in tracking metal dispersal in
\textsc{1sn}), we assumed that the metal abundances were homogeneous
within each supernova's ejecta.  The tracking of ejecta 
was carried out with Lagrangian passive tracer 
particles, $N_{\rm trace}=10^7$ in \textsc{1sn} and $(2-3)\times10^6$ per
supernova or a total of $N_{\rm trace}\approx 2\times10^7$ in \textsc{7sn}.  The tracers
allowed us to connect ejecta fluid elements to their origin in the explosions.
In \textsc{1sn} they allowed us to distinguish between
the ejecta originating in distinct mass shells within the explosion, and in \textsc{7sn} between the ejecta originating in
different supernovae.  It is worth
noting that the numerical limitations affecting tracers, e.g., associated with the finiteness of the order of the velocity field representation on the computational mesh, are distinct from the numerical teleportation problem mentioned in Section \ref{sec:thermodynamic}. Thus, the simulated metallicity is generally to be trusted in cells in which the Eulerian and Lagrangian tracers indicate consistent values of the metallicity.

\begin{figure*}
\begin{center}
\includegraphics[width=0.47\textwidth]{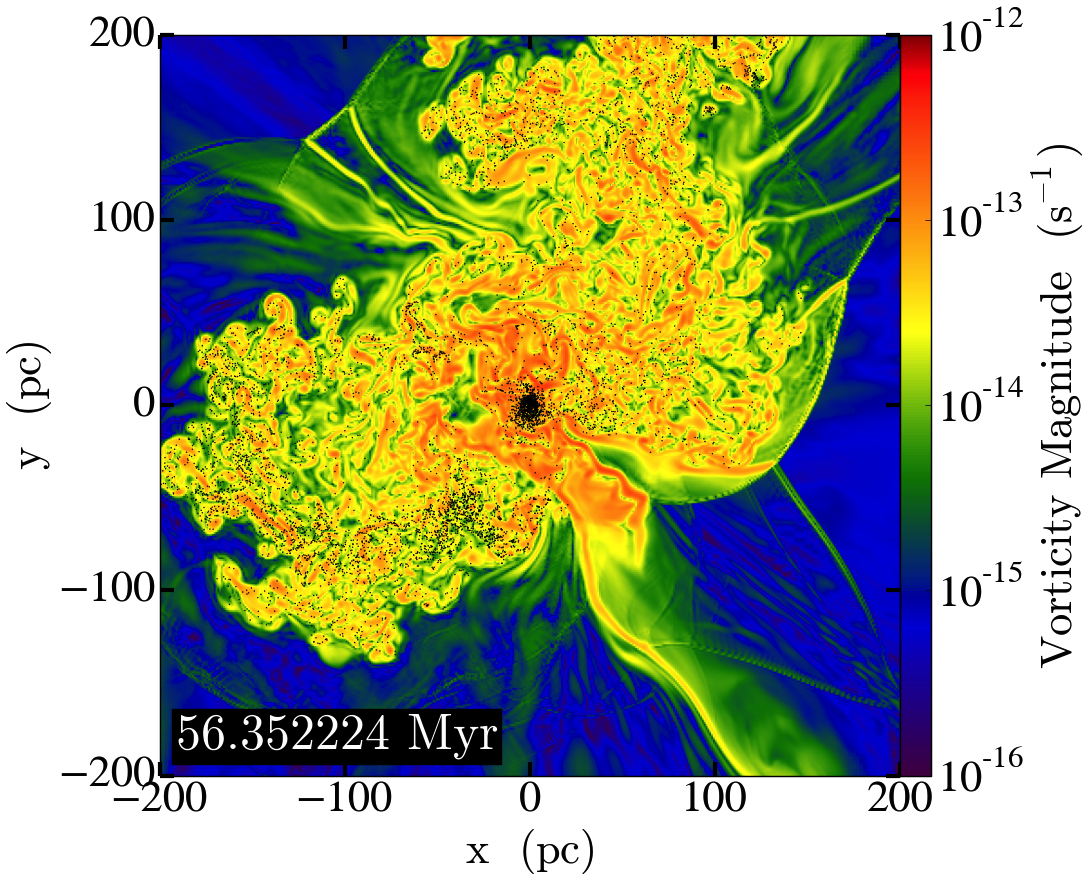}\hspace{0.5cm}
\includegraphics[width=0.47\textwidth]{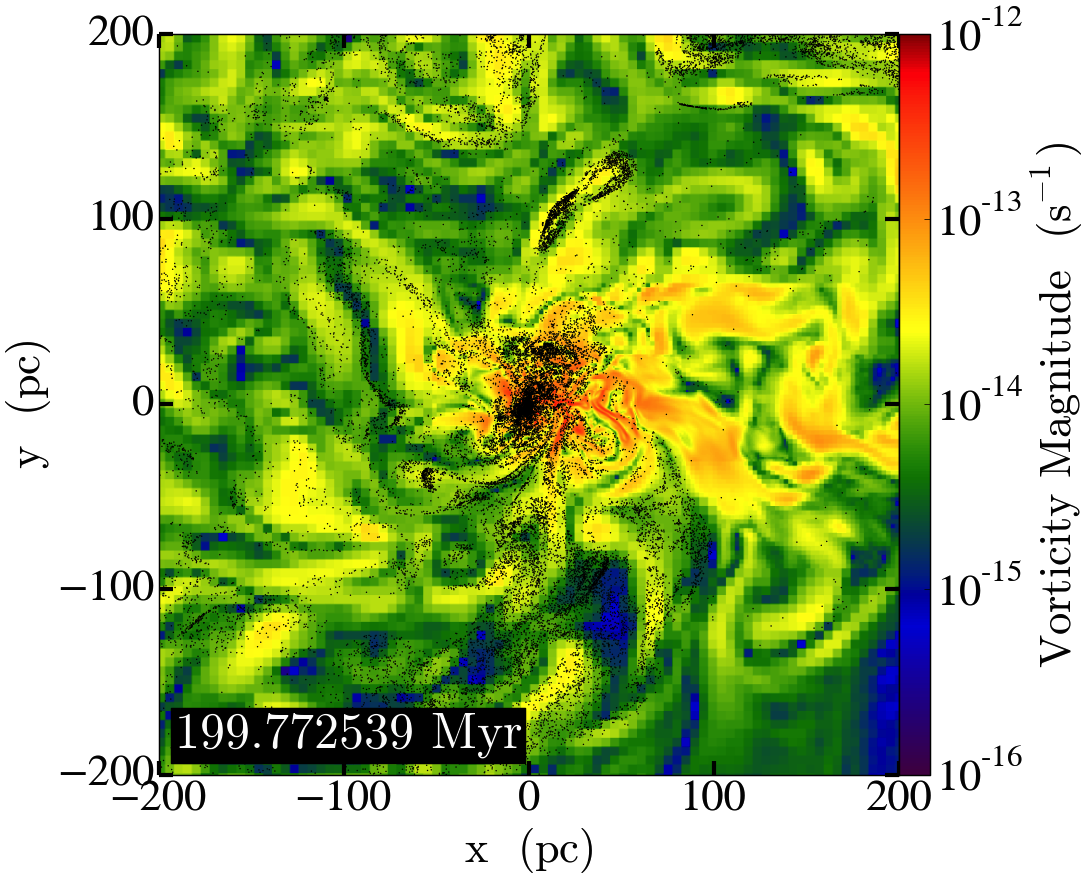}
\caption{A slice of vorticity magnitude $|\nabla\times\mathbf{v}|$ in the center
  of \textsc{1sn} (left) and \textsc{7sn} (right) overplotting ejecta tracer particles in a $5\,\textrm{pc}$
  thick slab containing the slice (black dots).  Note
  that metal ejecta are unmixed outside the central high-vorticity
  core, tens of parsecs in radius.\label{fig:vorticity}}
\end{center}
\end{figure*}

\section{Hydrodynamic Evolution}
\label{sec:results}

The supernova remnants evolved through the free expansion,
Sedov-Taylor, pressure-driven and momentum-conserving snowplow phases, and eventually, underwent
partial collapse along the direction of the cosmic gas inflow,
parallel to the filaments of the cosmic web.  In
\textsc{7sn}, the later exploding supernovae
expanded in the hot, low-density bubble evacuated
by earlier ones, with blastwaves colliding with the dense
radiative shell before the reverse shocks have completed traversal of
the ejecta.

For the hydrodynamic and chemical evolution of the
entire remnant or bubble, it is important that the photoionization by the
supernova progenitor stars did not completely photoevaporate the
densest primordial clouds inside the halo
\citep[][]{Abel:07,BlandHawthorn:11,Ritter:12}.  While the
photoevaporation flows are difficult to adequately resolve in
cosmological simulations, survival of
neutral clouds inside the primordial H\,II region is expected
from the analytical evaporation solutions of \citet{Bertoldi:90}.  The
densest clouds with central densities $n\lesssim 10\,\textrm{cm}^{-3}$ and
distances $\sim 50-100\,\textrm{pc}$ from the center of the halo
are associated with the filamentary inflow from the
cosmic web.  Supernova blastwaves swept past these clouds, partially
ablating them and depositing some ejecta material at the perimeters of
the clouds.  The blastwave-cloud interaction drove turbulence
inside the bubble.  The ablated primordial gas found itself inside the
hot, turbulent interior, where it appeared to be susceptible to
turbulent-stirring-aided mixing, especially in the multi-supernova simulation.  This
intra-bubble mixing resulted in a modest, factor of $\lesssim 10$ dilution of the
ejecta by the primordial gas.

After $\sim 0.13$ and $\sim1\,\textrm{Myr}$ from the (first) explosion
in simulation \textsc{1sn} and \textsc{7sn}, respectively, the ejecta material started
to accumulate in the pressure-driven snowplow shell. The shell was
thin $<10\,\textrm{pc}$ and not adequately resolved at grid resolution
$\sim0.5-1\,\textrm{pc}$. The insufficient resolution blurred the contact discontinuity and its
associated compositional gradient. As a result, the metallicity
derived from the passive mass scalar was diluted to
$Z\sim10^{-4}-10^{-3}$ in the thin shell.  Rayleigh-Taylor (RT)
fingers first
became prominent in the shell after $\sim 3\,\textrm{Myr}$ and 
then became extended, with length scales comparable to the radius of the
bubble, at $\sim20\,\textrm{Myr}$. The long-term hydrodynamical
evolution of the supernova bubble was highly anisotropic, with a
fraction of the ejecta and swept-up primordial medium
traveling many halo virial radii perpendicular to the cosmic web
filaments, and another $\sim50\%$ of the ejecta remaining within the virial
radius.\footnote{The morphological evolution of the metal-enriched volume bears similarities to both the centered and off-center idealized non-cosmological simulations of \citet{Webster:14}.}

After the supernova bubbles started to collapse, the bubble interiors
cooled to $\sim10^4\,\textrm{K}$ and began intermixing with the ambient
unshocked gas.  
Dual inflows fed toward the halo center: from the infall of unenriched,
primordial clouds (including from merging halos in
\textsc{7sn}),
and from the fragments of the buckling thin shell.  The
terminus of the inflows was a turbulent quasi-virialized (or quasi-hydrostatic) cloud in which
turbulence was stirred by gravitational infall.  
Figure \ref{fig:7sn_projections_slices} illustrates the overall
geometry of the metal distribution at this stage.  It shows a homogenized
low-metallicity interior surrounded by more metal rich, inhomogeneous
clouds.

\begin{figure*}
\begin{center}
\includegraphics[width=0.395\textwidth]{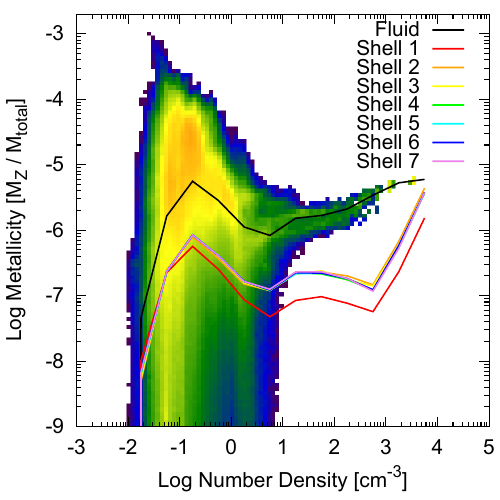}\hspace{1.0cm}
\includegraphics[width=0.395\textwidth]{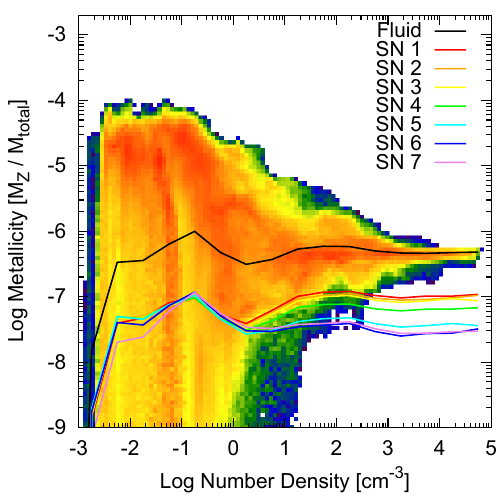}
\end{center}
\caption{Metallicity as a function of gas density at the end of the simulation
  \textsc{1sn} at $56\,\textrm{Myr}$ after the first explosion (left
  plot) and \textsc{7sn}  at $200\,\textrm{Myr}$ after the first explosion  (right plot).
From red to blue, the color scales with the logarithm
  of the fluid
  mass in the density and metallicity bin.  Solid curve is the mean
  metallicity and colored curves are the fractional contributions
  from the seven radial mass bins (\textsc{1sn}) and seven supernovae (\textsc{7sn}).\label{fig:phase_plot_metal_density}}
\end{figure*}

Figure \ref{fig:vorticity}, plotting the amplitude of the fluid vorticity,
shows that vortical time scales in the quasi-virialized cloud are
$|\nabla\times{\mathbf v}|^{-1} \sim 0.1-1\,\textrm{Myr}$, short
enough to facilitate turbulent-cascade-aided fluid mixing in $\sim10\,\mathrm{Myr}$. Vorticity is the highest near the center of each plot, where the
metal-enriched gas has gone into runaway gravitational collapse.
We expect the collapse to ultimately lead to the formation of second-generation
stars. 
Outside the quasi-virialized cloud, the vortical time scales are longer,
$10-100\,\textrm{Myr}$, precluding mixing.
Figure
\ref{fig:phase_plot_metal_density} shows the joint distribution of gas
density and metallicity in the two simulations.  Metallicity spread in
low-density gas is high, indicating a high degree of inhomogeneity.
The spread decreases
with increasing density, becoming narrow for $n\gtrsim10-
100\,\textrm{cm}^{-3}$.  The narrowing of the metallicity spread is a
consequence of rapid turbulent homogenization.   The densest gas has
negligible metallicity spread with $Z\approx5\times10^{-6}$ in
\textsc{1sn} and $5\times 10^{-7}$ in \textsc{7sn}.  We can conclude that 
the very first generation of metal-enriched stars will be chemically
homogeneous on the scale of a stellar group or cluster.  Recently, \citet{Feng:14} observed the same in simulations of star clusters forming in a very different environment, the Milky Way disk \citep[see, also,][]{BlandHawthorn:10}.
We shall see in the following section, however, that stellar abundance patterns 
in the first metal-enriched star forming systems will not
be simple superpositions of the yield patterns of the contributing supernovae.

\section{Monolithic and Biased Enrichment}

We procede to analyze how hydrodynamic dispersal depends on the
nucleosynthetic site.  
In analyzing simulation \textsc{1sn}, we
split the ejecta at supernova insertion into $N_{\rm bin}=7$ radial
bins containing equal ejecta masses and treat the bin index
$i=1,...,N_{\rm bin}$ as a crude proxy for the isotopic group synthesized in
the corresponding bin.  For example, the innermost bins could contain
the explosively synthesized
Fe peak and $\alpha$ elements, the intermediate bins could be rich in light
hydrostatic elements (C and O), and the outermost bins would contain H
and He.  This idealizes the explosion as preserving spherical
symmetry, which is certainly not the case, as symmetry is
strongly broken by convection preceding and during the collapse and by
RT fingering during the explosion (our simulations can
capture the RT instability in the remnant but not in the
explosion). However, we expect that the
radial ``dredging'' of elements by instabilities \citep[and rotation-driven
mixing, see, e.g.,][]{Meader:12} is incomplete and
that some radial stratification is preserved until the ejecta enter free expansion.
In analyzing simulation \textsc{7sn}, we ignore the stratification
inside each explosion and take the bin index $i=1,...,N_{\rm bin}$
to range over the $N_{\rm bin}=N_{\rm SN}=7$ supernovae.

We first examine the degree to which the ejecta in different
bins are stirred with each other (to be further discussed in a forthcoming
follow-up paper). In
\textsc{1sn}, the 
outer, high-velocity mass shells of the ejecta are well-stirred
between themselves, but
inner mass shells remain highly inhomogeneous.  In \textsc{7sn}, the ejecta from
supernovae separated by short time intervals are well-stirred with
each other.  The ejecta from later
supernovae, separated from the earlier supernovae by the longest time intervals, remain inhomogeneous.  

We call the enrichment by the source (a single supernova in
\textsc{1sn} and a cluster of supernovae in \textsc{7sn})
\emph{monolithic} if the density of isotope $A$ at
location $\mathbf{x}$ is proportional to the sum of the isotopic
yields of individual bins $Y_{i,A}$ over the bin index. Specifically,
\beq
\label{eq:monolithic}
\rho_A(\mathbf{x}) = Z(\mathbf{x}) \rho(\mathbf{x})  \frac{Y_{A}}
{\sum_A Y_A} , \ \  \ \ 
(\textrm{monolithic}) ,
\eeq
where 
\beq
Y_{A} \equiv \sum_{i=1}^{N_{\rm bin}} Y_{i,A}
\eeq
and $\rho(\mathbf{x})$ and $Z(\mathbf{x})$ are the total mass density
and total metallicity from the enrichment event, respectively, such 
that $\int Z \rho d^3x = \sum_A Y_A$.  It is
standard to assume \citep[e.g.,][]{Ting:12} that enrichment is indeed
monolithic so that each enrichment event $\mathcal{E}$ defines a unique chemical space vector
${\mathbf Y}^{(\mathcal{E})} \equiv (Y_{A_1}^{(\mathcal{E})},
Y_{A_2}^{(\mathcal{E})}, ...)\sim {\mathbf Y}^{(\mathcal{C})} $ representative of a class $\mathcal{C}$
of
nucleosynthetic sources (e.g., core collapse events with or without
r-process).  

If star formation takes place at
space-time points $(\mathbf{x}_k,t_k)$,
then stellar abundances are given by\footnote{Here and in what
  follows, coarse graining
  of $\rho_A$ on the spatial scales of star-forming clumps is implied.}
\beq
Z_{k,A} =\frac{1}{\rho (\mathbf{x}_k,t_k)} \sum_{\mathcal{E}}\rho_A
^{(\mathcal{E})} (\mathbf{x}_k,t_k), \ \ \ 1\leq k\leq N_{\rm star}
\eeq
The goal of chemical abundance analysis is then to isolate
nucleosynthetic source classes by recovering their yield vectors
$\mathbf{Y}^{(\mathcal{C})}$ from stellar abundance data $Z_{k,A}$.
The dimensionality of the chemical abundance space corresponds to the
number of different nucleosynthetic classes.

The metallicities resulting from individual events $Z^{(\mathcal{E})}(\mathbf{x},t)$
are random variables determined by the hydrodynamics of metal
dispersal. They endow 
stellar metallicities with scatter. If a sufficient number of events
contributes and the events are statistically independent, 
the central limit theorem implies that the scatter is Gaussian,
justifying the PCA approach.

It cannot be taken for granted, however, that the source contributions
are monolithic (Eqn.~\ref{eq:monolithic}), because nucleosynthetic products are 
injected into the hydrodynamic environment with different velocities
and at different times and are transported differently to their star formation
sites.  In perfect generality, for each nucleosynthetic event,
\bea
\rho_A(\mathbf{x}) &=& \sum_{i=1}^{N_{\rm bin}} Z_i(\mathbf{x}) \rho(\mathbf{x})  \frac{Y_{i,A}}
{\sum_A Y_{i,A}} \nonumber\\ &=&  Z(\mathbf{x}) \rho(\mathbf{x})
\frac{\sum_{i=1}^{N_{\rm bin}} w_i(\mathbf{x}) Y_{i,A}}
{\sum_A Y_A} ,
\eea
where $Z_i(\mathbf{x})$ is the total metallicity at $\mathbf{x}$ due to enrichment by
bin $i$ such that $\int Z_i \rho d^3 x= \sum_A Y_{i,A}$, and in the
second line, we introduced the weights 
\beq
w_i(\mathbf{x})
= \frac{Z_i(\mathbf{x})}{Z(\mathbf{x})}\, \frac{ \sum_A Y_A}{\sum_A
Y_{i,A} } 
\eeq
for straightforward comparison with the monolithic case
(Eq.~\ref{eq:monolithic}).

If source contributions are monolithic then $w_i \equiv 1/N_{\rm bin}$,
but in general, the enrichment at a specific location could be
biased toward some bins, giving them higher weights.  The
dimensionality of the chemical abundance space can now be much larger
than the number of nucleosynthetic source classes.
The weights
encapsulate the biases that hydrodynamics introduces into abundance
patterns.
If the weights are completely random and possess unknown statistics,
this introduces uncontrolled biases frustrating the
recovery of nucleosynthetic source classes from stellar abundance
data.  Here, we make the first step toward characterizing the nature
of the biases, aspiring to detect regularities that can be factored
into chemical abundance analysis.  Figure
\ref{fig:phase_plot_metal_density} shows 
bin-specific metallicities $Z_i$ spherically averaged around the
center of gravitational collapse.  In both simulations, in dense gas $n\gtrsim 1\,\textrm{cm}^{-3}$, 
departures from monolithic enrichment are evident.  In \textsc{1sn},
the innermost radial bin, which is expected to carry explosive elements, is
deficient by a factor of $\sim 3$ relative to the other bins, $\langle
w_1\rangle\sim
\frac{1}{3}\langle w_2\rangle$ and $\langle w_2\rangle \sim...\sim
\langle w_7\rangle$, where the averages refer to gas-mass-weighted
averages of $w_i$ in the dense gas.  In
\textsc{7sn}, ejecta from the first two supernovae are $\sim 3$ times as abundant as the
ejecta from the last two supernovae, $\langle w_1\rangle \sim \langle
w_2\rangle  \sim 3\langle w_6\rangle \sim 3\langle w_7\rangle$.  

We propose the following physical interpretation.  In the case of the
solitary supernova, the reverse shock raises the inner ejecta shells,
which it sweeps later and at lower densities, to a higher entropy than the outer shells, which
it sweeps earlier and at higher densities.  This can be seen by
considering the Sedov-Taylor point
explosion, in which the pressure asymptotes to a constant value near the center, but
density decreases toward the center 
as $\rho\propto r^{3/(\gamma-1)}$.  Therefore, with $\gamma=5/3$, the entropy $s=\ln(
P/\rho^\gamma)+{\rm const}$ rises toward the
center of the ejecta as $s\sim -\frac{27}{10}\ln r+\textrm{const}$.
This is significant because as the ejecta become quasi-isobaric inside the
remnant, the radiative cooling time is a steeply increasing function
of entropy.  The outer ejecta shells cool first and are incorporated
into the snowplow shell, while the innermost ejecta avoid cooling.
As the remnant stalls and collapses, the innermost ejecta, having higher
entropy, are outward buoyant and rise to larger radii, thus avoiding
collapse and allowing
low-entropy outer ejecta mass shells to fall in to enrich the
quasi-virialized cloud.  In \textsc{1sn}, the ejecta tracer
particle radii containing $25\%$ of the ejecta in bins $1$ (the
innermost bin) and $2$ cross at $\sim 0.2\,\textrm{Myr}$ after the explosion
and then bin $1$ interchanges with bins $\geq 3$ at $\sim
0.3\,\textrm{Myr}$.  

In the case of clustered supernovae, the situation is similar, but now
since the density inside the supernova bubble keeps dropping as the
bubble expands, the ejecta from later supernovae are on average raised to higher
entropies than those of earlier ones and are less susceptible to
cooling.  Since the ejecta of the latest supernovae remain hot, they
are outward buoyant. Upon the collapse of the bubble, the hot ejecta
of the later supernovae interchange with the cooled ejecta
of the earlier supernovae.  The latter fall in to enrich the central cloud.

\section{Confronting the Empirical Record}
\label{sec:discussion}

Our results have important implications for understanding the
early stages of cosmic chemical evolution. More precisely, the
hydrodynamic biases in the transport of individual elements, introduced by
the post-explosion evolution, need to be taken into account when 
confronting the empirical abundance trends. The ultimate goal here is
to achieve a robust mapping from the observed abundance pattern in
metal-poor stars or systems thereof to the individual sources of 
those metals. In the absence of any monolithic mapping between
sources and fossil record (see Section~4), the hydrodynamic
transport process constitutes the missing link in our current understanding.
Simulations along the lines of our exploratory work here promise to
bridge this crucial gap. However, we can already now address a long-standing
problem in Galactic chemical evolution in a new light. This concerns
the prevalence of peculiar abundance ratios in low-metallicity stars and
systems.

To briefly summarize the main phenomenology, observations of
metal-poor stars in the Galactic halo, assembled over more than two
decades, have provided intriguing constraints on the nature of early
chemical evolution \cite[reviewed in][]{Beers:05,Frebel:13a,Karlsson:13}. The
current state of the art is defined by a large sample of halo red giant stars,
where key lines are sufficiently strong to enable high signal-to-noise
spectroscopy \citep{Cayrel:04,Francois:07}. The main lessons are
two-fold: both $\alpha$-elements (Mg, Ca, Si, Ti), and iron-peak elements (V to Zn)
exhibit extremely small scatter, down to $\textrm{[Fe/H]}\sim -3.5$. The neutron-capture
elements, comprising elements beyond Zn, on the other hand, exhibit equally small 
scatter down to $\textrm{[Fe/H]}\sim -3$, but show extremely large scatter, up to 5\,dex, below this
\citep{Qian:02,Truran:02,Sneden:08}. Finally, the lighter elements (C, N, O)
again show large abundance variations at $\textrm{[Fe/H]}< -4$, and approach well-defined
trends for less metal-poor stars.\footnote{The skewing of abundance patterns by post-supernova hydrodynamics may also be the explanation of abundance anomalies \citep[e.g.,][]{Feltzing:09,Cohen:13,Yong:13a} found in primitive stellar systems in the local universe.}
The most dramatic manifestation of this transition to well-behaved
abundance trends, once a threshold metallicity is reached, is provided by
the huge scatter in r-process abundances,
seen in Galactic halo stars with $\textrm{[Fe/H]}\lesssim -3$
\citep[reviewed in][]{Sneden:08}. The origin of r-process nucleosynthesis,
such as the mass and properties of the supernova progenitor star, is still
highly uncertain. There is, however, tentative evidence that the r-process
might operate in progenitor stars with a very narrow mass range, possibly
close to the lower-mass limit for core-collapse supernovae
\citep[e.g.,][]{Qian:08}.

It has been challenging to explain all of these
trends within one comprehensive framework, but our work suggests
a promising {\it Ansatz} to do so. Basically, our results show that
early enrichment is differential, non-monolithic in nature. This specifically
implies that second-generation star formation does not sample the enrichment
from the full IMF, and possibly not even that from a single explosion (see Fig.~3).
Our simulations build on earlier analytical work that had postulated 
a minimum number of supernovae, $N_{\rm SN}\gtrsim20$, needed to average out
any yield inhomogeneity from individual explosion sites \citep[see, e.g.,][]{Tsujimoto:98,Tsujimoto:99}.
This absence of effective source-averaging, then, such that only a small number  
of explosion sites contribute, is the key requirement to preserve peculiar
abundance ratios. Next to the classical r-process elements, a similar
explanation may pertain to
the strong odd-even pattern that is predicted for pair-instability supernova
enrichment, but has not been detected so far \citep{Heger:02,Karlsson:08}.
We note that our explanation of the r-process record, understanding huge
scatter as a result of sparse IMF-sampling, where only a very narrow progenitor mass
range gives rise to the r-process, does not require any stochastic contribution
from neutron star mergers, as has recently been suggested \citep{Shen:14,Voort:14}.

Once cosmological structure formation advances to more massive systems, with
deeper potential wells to facilitate the near-uniform mixing of the ejecta 
from a large number of supernovae, convergence towards well-defined, smooth
abundance trends will set in. Extragalactic observations targeting systems
of greatly different virial mass are in agreement with this overall
picture. Specifically, recent
medium- and high-resolution spectroscopy of red giant stars in Milky Way
dwarf satellites has established that their abundance properties, including
the degree of scatter, are indistinguishable from the metal-poor tail of the
Galactic halo stars \citep{Frebel:10b,Frebel:14}. A complementary view into 
early metal enrichment is provided by the abundances measured in damped
Lyman-$\alpha$ (DLA) systems. Here, the evidence points towards extremely low
scatter, at least for the prominent $\alpha$-elements \citep{Cooke:11b,Becker:12}. This
may be indicative of the onset of efficient gas-phase mixing in the deep potential
wells of the DLA dark matter host halos.
An intriguing question for future simulation work
is to test whether the empirical threshold metallicity, roughly measured by [Fe/H], for
the disappearance of such anomalous abundance signatures can be reproduced.

Of special interest is the observed dichotomy of carbon-enhanced and carbon-normal
metal-poor stars \citep[e.g.,][]{Beers:05,Gilmore:13}. 
\citet[][hereafter CM14]{Cooke:14} have recently suggested that the strength of supernova driven
outflows may be responsible for this bimodality.
These authors 
invoke two classes of supernovae with greatly different times for recovery
from the supernova explosions with different explosion energies.  More
energetic explosions $E_{\rm SN}\gg 10^{51}\,\textrm{erg}$ imply
longer recovery times \citep{Jeon:14}.  
According to CM14, 
rapid recovery is connected to weak explosions with large carbon
overabundances, 
and slow recovery with strong explosions with normal carbon yields.
Our simulation \textsc{1sn}, where we find the return of supernova
ejecta into the halo center deficient in the
ejecta originating from the innermost $10\%$ of the ejecta, 
suggests another mechanism for  carbon-enhanced metal poor (CEMP)
stars.  
A core-collapse explosion ejects a normal carbon-to-iron ratio $[{\rm
  C}/{\rm Fe}]< 1$, 
but the iron-bearing ejecta are
raised to a higher entropy upon reverse shock traversal than the
carbon-bearing ejecta, and subsequently do not cool and settle
into the halo center to form second-generation stars.

\section{Summary and Conclusions}
\label{sec:conclusions}

We have carried out two complementary very-high-resolution cosmological simulations of how
the metals produced in the first supernova explosions are transported
into the cold, dense gas out of which the second-generation of (Population~II)
stars is formed. The first simulation followed the ejecta from a single
explosion, whereas the second traced the metal dispersal from seven sources.
We arrived at two main conclusions. The re-condensed Population~II star forming material
exhibits strong turbulent vorticity, implying a likely fine-grained turbulent
mixing of gas down to very small, unresolved, scales. Stellar clusters or
groups forming out of this material are thus predicted to be chemically 
uniform, unless any self-enrichment process may operate during the
later stages of stellar evolution. Our results also indicate
that the hydrodynamic metal transport proceeds differentially, such that 
the monolithic mapping of source abundances into the fossil record is broken.

The hydrodynamically biased nature of early metal enrichment, as demonstrated
in our pathfinder simulations, has multiple implications, requiring
that we rethink a number of our traditional assumptions and methodologies.
On the theory side, a common approach to chemical enrichment is to
assume that homogenization of the chemical composition is
instantaneous and complete in certain ``mixing
volumes,'' normally centered on nucleosynthetic sources, but that the
volumes themselves occur stochastically and intermittently, tracing
star formation.  This approach provides a rudimentary model
of inhomogeneous chemical evolution, but since the choice of mixing
volumes is ad hoc, its predictive power is limited.  
It is standard to motivate the choice of mixing volumes by considering the
spatial extent of supernova remnants and galactic superbubbles,
assuming that the medium is chemically homogeneous within these structures 
\citep{Argast:00,Oey:00,Oey:03,Karlsson:05a,Karlsson:05b,Karlsson:08,BlandHawthorn:10,Corlies:13}.
In the absence of monolithic source mapping, it is not obvious how to
adjust this technique.

Our results also present
a challenge to the standard interpretational framework of near-field cosmology.
The hydrodynamic biases need to be quantified as a function of the star forming environment by carrying out a dedicated program of simulations.
Once these chemical transport `maps' are in hand, the full power of 
stellar archaeology can be unleashed. Such a program is very timely, given
the advent of large survey projects, including the dedicated efforts connected
to Gaia-ESO, which promise a record of early 
chemical evolution in unprecedented detail.

\section*{Acknowledgments}

We acknowledge helpful conversations with Anna Frebel, John Scalo, and Craig Wheeler.
The \textsc{flash} code was in part developed by the DOE-supported Flash Center
for Computational Science at the University of Chicago. The authors
acknowledge the Texas Advanced Computing Center at The University of
Texas at Austin for providing HPC resources under XSEDE allocation
TG-AST120024. CSS is grateful for support provided by the NASA Earth
and Space Science Fellowship (NESSF) program. This study was supported
by the NSF grant AST-1009928 and by the NASA grant NNX09AJ33G. Many of the
visualizations were made with the \textsc{yt} package \citep{Turk:11}.

\footnotesize{

}

\label{lastpage}


\begin{thebibliography}{}


%The H II Region of a Primordial Star
\bibitem[Abel et al.(2007)]{Abel:07} Abel, T., Wise, J.~H., 
\& Bryan, G.~L.\ 2007, \apjl, 659, L87 

%Metal-poor halo stars as tracers of ISM mixing processes during halo formation
\bibitem[Argast et 
al.(2000)]{Argast:00} Argast, D., Samland, M., Gerhard, O.~E., \& Thielemann, F.-K.\ 2000, \aap, 356, 873 

%Toward Realistic Progenitors of Core-collapse Supernovae
\bibitem[Arnett 
\& Meakin(2011)]{Arnett:11} Arnett, W.~D., \& Meakin, C.\ 2011, \apj, 733, 78 

%The First Generation of Stars: First Steps toward Chemical Evolution of Galaxies
\bibitem[Audouze 
\& Silk(1995)]{Audouze:95} Audouze, J., \& Silk, J.\ 1995, \apjl, 451,
L49 

%Iron and α-element Production in the First One Billion Years after the Big Bang
\bibitem[Becker et al.(2012)]{Becker:12} Becker, G.~D., Sargent, 
W.~L.~W., Rauch, M., \& Carswell, R.~F.\ 2012, \apj, 744, 91 

%The Discovery and Analysis of Very Metal-Poor Stars in the Galaxy
\bibitem[Beers 
\& Christlieb(2005)]{Beers:05} Beers, T.~C., \& Christlieb, N.\ 2005, \araa, 43, 531 

%The photoevaporation of interstellar clouds. II - Equilibrium
%cometary clouds
\bibitem[Bertoldi 
\& McKee(1990)]{Bertoldi:90} Bertoldi, F., \& McKee, C.~F.\ 1990, \apj, 354, 529 

\bibitem[Bertschinger(2001)]{Bertschinger:01} Bertschinger, E.\ 2001, 
\apjs, 137, 1 

%The Chemical Signatures of the First Star Clusters in the Universe
\bibitem[Bland-Hawthorn et al.(2010)]{BlandHawthorn:10} Bland-Hawthorn, 
J., Karlsson, T., Sharma, S., Krumholz, M., 
\& Silk, J.\ 2010, \apj, 721, 582 

%The Minimum Mass for a Dwarf Galaxy
\bibitem[Bland-Hawthorn et al.(2011)]{BlandHawthorn:11} Bland-Hawthorn, 
J., Sutherland, R., \& Karlsson, T.\ 2011, EAS Publications Series, 48, 397

%Pre-Reionization Fossils, Ultra-Faint Dwarfs, and the Missing Galactic Satellite Problem
\bibitem[Bovill 
\& Ricotti(2009)]{Bovill:09} Bovill, M.~S., \& Ricotti, M.\ 2009, \apj, 693, 1859 

%The Primeval Populations of the Ultra-faint Dwarf Galaxies
\bibitem[Brown et al.(2012)]{Brown:12} Brown, T.~M., Tumlinson, 
J., Geha, M., et al.\ 2012, \apjl, 753, L21 

%First stars V - Abundance patterns from C to Zn and supernova yields in the early Galaxy
\bibitem[Cayrel et 
al.(2004)]{Cayrel:04} Cayrel, R., Depagne, E., Spite, M., et al.\ 2004, \aap, 416, 1117 

%The Formation and Fragmentation of Disks Around Primordial Protostars
\bibitem[Clark et al.(2011a)]{Clark:11a} Clark, P.~C., Glover, 
S.~C.~O., Smith, R.~J., et al.\ 2011a, Science, 331, 1040 

%Gravitational Fragmentation in Turbulent Primordial Gas and the
%Initial Mass Function of Population III Stars
\bibitem[Clark et al.(2011b)]{Clark:11b} Clark, P.~C., Glover, 
S.~C.~O., Klessen, R.~S., \& Bromm, V.\ 2011b, \apj, 727, 110 

%Normal and Outlying Populations of the Milky Way Stellar Halo at [Fe/H] <-2
\bibitem[Cohen et al.(2013)]{Cohen:13} Cohen, J.~G., Christlieb, 
N., Thompson, I., et al.\ 2013, \apj, 778, 56 



%DONTCITE: Becker criticism. See Frebel & Norris 2015, section 6.2
%A carbon-enhanced metal-poor damped Lyα system: probing gas from
%Population III nucleosynthesis?
%\bibitem[Cooke et al.(2011)]{Cooke:11a} Cooke, R., Pettini, M., 
%Steidel, C.~C., Rudie, G.~C., \& Jorgenson, R.~A.\ 2011, \mnras, 412, 1047 

%The most metal-poor damped Lyα systems: insights into chemical evolution in the very metal-poor regime
\bibitem[Cooke et al.(2011)]{Cooke:11b} Cooke, R., Pettini, M., 
Steidel, C.~C., Rudie, G.~C., \& Nissen, P.~E.\ 2011, \mnras, 417, 1534 

%Carbon-Enhanced Metal-Poor Stars: Relics from the Dark Ages
\bibitem[Cooke 
\& Madau(2014)]{Cooke:14} Cooke, R., \& Madau, P.\ 2014, arXiv:1405.7369 

%Chemical Abundance Patterns and the Early Environment of Dwarf
%Galaxies
\bibitem[Corlies et al.(2013)]{Corlies:13} Corlies, L., Johnston, 
K.~V., Tumlinson, J., \& Bryan, G.\ 2013, \apj, 773, 105 

%The Three Dimensional Evolution to Core Collapse of a Massive Star
\bibitem[Couch et al.(2015)]{Couch:15} Couch, S.~M., 
Chatzopoulos, E., Arnett, W.~D., \& Timmes, F.~X.\ 2015, arXiv:1503.02199 

%The evaporation of spherical clouds in a hot gas. I - Classical and saturated mass loss rates
\bibitem[Cowie \& McKee(1977)]{Cowie:77} Cowie, L.~L., \& McKee, C.~F.\ 1977, \apj, 211, 135 




%DONTCITE
%Stellar Archaeology in the Galactic Halo with the Ultra-faint Dwarfs. VI. Ursa Major II
%\bibitem[Dall'Ora et al.(2012)]{DallOra:12} Dall'Ora, M., 
%Kinemuchi, K., Ripepi, V., et al.\ 2012, \apj, 752, 42

%DONTCITE
%Mixing Timescales in a Supernova-driven Interstellar Medium
%\bibitem[de Avillez 
%\& Mac Low(2002)]{deAvillez:02} de Avillez, M.~A., \& Mac Low, M.-M.\ 2002, \apj, 581, 1047 

%The GALAH Survey: Scientific Motivation
\bibitem[De Silva et al.(2015)]{DeSilva:15} De Silva, G.~M., 
Freeman, K.~C., Bland-Hawthorn, J., et al.\ 2015, arXiv:1502.04767 

\bibitem[Draine(2011)]{Draine:11} Draine, B.~T.\ 2011, Physics of 
the Interstellar and Intergalactic Medium by Bruce T.~Draine.~Princeton 
University Press, 2011.~ISBN: 978-0-691-12214-4,  



%A Case Study of Small-scale Structure Formation in Three-dimensional Supernova Simulations
\bibitem[Ellinger et al.(2012)]{Ellinger:12} Ellinger, C.~I., 
Young, P.~A., Fryer, C.~L., \& Rockefeller, G.\ 2012, \apj, 755, 160

%First Simulations of Core- Collapse Supernovae to Supernova Remnants with SNSPH
%\bibitem[Ellinger et al.(2013)]{Ellinger:13} Ellinger, C.~I., 
%Rockefeller, G., Fryer, C.~L., Young, P.~A., \& Park, S.\ 2013, arXiv:1305.4137 



%Evidence of enrichment by individual SN from elemental abundance
%ratios in the very metal-poor dSph galaxy Boötes I
\bibitem[Feltzing et 
al.(2009)]{Feltzing:09} Feltzing, S., Eriksson, K., Kleyna, J., \& Wilkinson, M.~I.\ 2009, \aap, 508, L1 



%Early turbulent mixing as the origin of chemical homogeneity in open star clusters
\bibitem[Feng \& Krumholz(2014)]{Feng:14} Feng, Y., \& Krumholz, M.~R.\ 2014, \nat, 513, 523 

%The 2013 Release of Cloudy
\bibitem[Ferland et al.(2013)]{Ferland:13} Ferland, G.~J., Porter, 
R.~L., van Hoof, P.~A.~M., et al.\ 2013, RMxAA, 49, 137 

%Mixing metals in the early Universe
%\bibitem[Ferrara et al.(2000)]{2000MNRAS.319..539F} Ferrara, A., Pettini, 
%M., \& Shchekinov, Y.\ 2000, \mnras, 319, 539 

%First stars. VIII. Enrichment of the neutron-capture elements in the early Galaxy
\bibitem[Fran{\c c}ois et 
al.(2007)]{Francois:07} Fran{\c c}ois, P., Depagne, E., Hill, V., et al.\ 2007, \aap, 476, 935 

%Nucleosynthetic signatures of the first stars
%\bibitem[Frebel et al.(2005)]{Frebel:05} Frebel, A., Aoki, W., 
%Christlieb, N., et al.\ 2005, \nat, 434, 871 

%Stellar archaeology: Exploring the Universe with metal-poor stars
%\bibitem[Frebel(2010)]{Frebel:10a} Frebel, A.\ 2010, Astronomische 
%Nachrichten, 331, 474 

%High-Resolution Spectroscopy of Extremely Metal-Poor Stars in the Least Evolved Galaxies: Ursa Major II and Coma Berenices
\bibitem[Frebel et al.(2010)]{Frebel:10b} Frebel, A., Simon, 
J.~D., Geha, M., \& Willman, B.\ 2010, \apj, 708, 560 

% Chemical Signatures of the First Galaxies: Criteria for One-shot Enrichment
\bibitem[Frebel 
\& Bromm(2012)]{Frebel:12} Frebel, A., \& Bromm, V.\ 2012, \apj, 759, 115 

%Metal-Poor Stars and the Chemical Enrichment of the Universe
\bibitem[Frebel 
\& Norris(2013)]{Frebel:13a} Frebel, A., \& Norris, J.~E.\ 2013, Planets, Stars and Stellar Systems.~Volume 5: Galactic Structure and Stellar Populations, 55 

%Exploring the Universe with Metal-Poor Stars
%\bibitem[Frebel(2013)]{Frebel:13b} Frebel, A.\ 2013, Astrophysics 
%and Space Science Library, 396, 377 

%CHEMOBS
%Segue 1: An Unevolved Fossil Galaxy from the Early Universe
\bibitem[Frebel et al.(2014)]{Frebel:14} Frebel, A., Simon, 
J.~D., \& Kirby, E.~N.\ 2014, \apj, 786, 74

%Near-Field Cosmology with Metal-Poor Stars
\bibitem[Frebel 
\& Norris(2015)]{Frebel:15} Frebel, A., \& Norris, J.~E.\ 2015, arXiv:1501.06921 



\bibitem[Fryxell et al.(2000)]{Fryxell:00} Fryxell, B., Olson, K., 
Ricker, P., et al.\ 2000, \apjs, 131, 273 

%The Gaia-ESO Public Spectroscopic Survey
\bibitem[Gilmore et al.(2012)]{Gilmore:12} Gilmore, G., Randich, 
S., Asplund, M., et al.\ 2012, The Messenger, 147, 25 


%Elemental Abundances and their Implications for the Chemical
%Enrichment of the Boötes I Ultrafaint Galaxy
\bibitem[Gilmore et al.(2013)]{Gilmore:13} Gilmore, G., Norris, 
J.~E., Monaco, L., et al.\ 2013, \apj, 763, 61 


%THE FIRST GALAXIES: CHEMICAL ENRICHMENT, MIXING, AND STAR FORMATION
\bibitem[Greif et al.(2010)]{Greif:10} Greif, T.~H., Glover, 
S.~C.~O., Bromm, V., \& Klessen, R.~S.\ 2010, \apj, 716, 510 

\bibitem[Glover \& Clark(2014)]{Glover:14} Glover, S.~C.~O., \& Clark, P.~C.\
2014, \mnras, 437, 9 

\bibitem[G{\'o}rski et al.(2005)]{Gorski:05} G{\'o}rski, K.~M., 
Hivon, E., Banday, A.~J., et al.\ 2005, \apj, 622, 759 

%Simulations on a Moving Mesh: The Clustered Formation of Population
%III Protostars
\bibitem[Greif et al.(2011)]{Greif:11} Greif, T.~H., Springel, 
V., White, S.~D.~M., et al.\ 2011, \apj, 737, 75 
	
%Formation and evolution of primordial protostellar systems
\bibitem[Greif et al.(2012)]{Greif:12} Greif, T.~H., Bromm, V., 
Clark, P.~C., et al.\ 2012, \mnras, 424, 399 

%DONTCITE
%HOW MANY NUCLEOSYNTHESIS PROCESSES EXIST AT LOW METALLICITY?
%\bibitem[Hansen et al.(2014)]{2014ApJ...797..123H} Hansen, C.~J., Montes, 
%F., \& Arcones, A.\ 2014, \apj, 797, 123 



%The Nucleosynthetic Signature of Population III
\bibitem[Heger 
\& Woosley(2002)]{Heger:02} Heger, A., \& Woosley, S.~E.\ 2002, \apj, 567, 532 

%Nucleosynthesis and Evolution of Massive Metal-free Stars
\bibitem[Heger 
\& Woosley(2010)]{Heger:10} Heger, A., \& Woosley, S.~E.\ 2010, \apj, 724, 341 




%Protostellar Feedback Halts the Growth of the First Stars in the Universe
\bibitem[Hosokawa et al.(2011)]{Hosokawa:11} Hosokawa, T., Omukai, 
K., Yoshida, N., \& Yorke, H.~W.\ 2011, Science, 334, 1250 

%DONTCITE
%CHEMOBS
%Chemical compositions of six metal-poor stars in the ultra-faint
%dwarf spheroidal galaxy Boötes I
%\bibitem[Ishigaki et 
%al.(2014)]{Ishigaki:14} Ishigaki, M.~N., Aoki, W., Arimoto, N., \& Okamoto, S.\ 2014, \aap, 562, A146 

%Recovery from population III supernova explosions and the onset of second generation star formation
\bibitem[Jeon et al.(2014)]{Jeon:14} Jeon, M., Pawlik, A.~H., 
Bromm, V., \& Milosavljevic, M.\ 2014, arXiv:1407.0034 

%DONTCITE
%The Chemical Imprint of Silicate Dust on the Most Metal-poor Stars
%\bibitem[Ji et al.(2014)]{Ji:14} Ji, A.~P., Frebel, A., 
%\& Bromm, V.\ 2014, \apj, 782, 95 

%Radiative shocks and hydrogen molecules in pregalactic gas - The effects of postshock radiation
\bibitem[Kang 
\& Shapiro(1992)]{Kang:92} Kang, H., \& Shapiro, P.~R.\ 1992, \apj, 386, 432 


%DONTCITE
%Chemical abundance patterns - fingerprints of nucleosynthesis in the first stars
%\bibitem[Karlsson 
%\& Gustafsson(2001)]{Karlsson:01} Karlsson, T., \& Gustafsson, B.\ 2001, \aap, 379, 461

%Stochastic chemical enrichment in metal-poor systems. I. Theory
\bibitem[Karlsson(2005)]{Karlsson:05a} Karlsson, T.\ 2005, \aap, 439, 93 

%Stochastic chemical enrichment in metal-poor systems. II. Abundance
%ratios and scatter
\bibitem[Karlsson 
\& Gustafsson(2005)]{Karlsson:05b} Karlsson, T., \& Gustafsson, B.\ 2005, \aap, 436, 879 

%DONTCITE
%Primordial Stellar Feedback and the Origin of Hyper-Metal-poor Stars
%\bibitem[Karlsson(2006)]{Karlsson:06} Karlsson, T.\ 2006, \apjl, 
%641, L41 

%Uncovering the Chemical Signature of the First Stars in the Universe
\bibitem[Karlsson et al.(2008)]{Karlsson:08} Karlsson, T., Johnson, 
J.~L., \& Bromm, V.\ 2008, \apj, 679, 6 

%Pregalactic metal enrichment: The chemical signatures of the first
%stars
\bibitem[Karlsson et al.(2013)]{Karlsson:13} Karlsson, T., Bromm, 
V., \& Bland-Hawthorn, J.\ 2013, Reviews of Modern Physics, 85, 809 

%Uncovering Extremely Metal-Poor Stars in the Milky Way's Ultrafaint Dwarf Spheroidal Satellite Galaxies
\bibitem[Kirby et al.(2008)]{Kirby:08} Kirby, E.~N., Simon, 
J.~D., Geha, M., Guhathakurta, P., \& Frebel, A.\ 2008, \apjl, 685, L43 

%DONTCITE
%Multi-element Abundance Measurements from Medium-resolution
%Spectra. III. Metallicity Distributions of Milky Way Dwarf Satellite
%Galaxies
%\bibitem[Kirby et al.(2011)]{Kirby:11} Kirby, E.~N., Lanfranchi, 
%G.~A., Simon, J.~D., Cohen, J.~G., \& Guhathakurta, P.\ 2011, \apj, 727, 78

%DONTCITE
%The Universal Stellar Mass-Stellar Metallicity Relation for Dwarf Galaxies
%\bibitem[Kirby et al.(2013)]{Kirby:13} Kirby, E.~N., Cohen, 
%J.~G., Guhathakurta, P., et al.\ 2013, \apj, 779, 102 

%DONTCITE
%The Origin of Low [α/Fe] Ratios in Extremely Metal-poor Stars
%\bibitem[Kobayashi et al.(2014)]{Kobayashi:14} Kobayashi, C., 
%Ishigaki, M.~N., Tominaga, N., \& Nomoto, K.\ 2014, \apjl, 785, L5 

%DONTCITE
%Complexity on Small Scales: The Metallicity Distribution of the Carina Dwarf Spheroidal Galaxy
%\bibitem[Koch et al.(2006)]{Koch:06} Koch, A., Grebel, E.~K., 
%Wyse, R.~F.~G., et al.\ 2006, \aj, 131, 895

%DONTCITE
%Complexity on Small Scales. II. Metallicities and Ages in the Leo II Dwarf Spheroidal Galaxy
%\bibitem[Koch et al.(2007)]{Koch:07} Koch, A., Grebel, E.~K., 
%Kleyna, J.~T., et al.\ 2007, \aj, 133, 270

%DONTCITE
%Complexity on Small Scales. III. Iron and α Element Abundances in the Carina Dwarf Spheroidal Galaxy
%\bibitem[Koch et al.(2008)]{Koch:08} Koch, A., Grebel, E.~K., 
%Gilmore, G.~F., et al.\ 2008, \aj, 135, 1580 


\bibitem[Komatsu et al.(2011)]{Komatsu:11} Komatsu, E., Smith, 
K.~M., Dunkley, J., et al.\ 2011, \apjs, 192, 18 

%DONTCITE
%The Origin of Carbon Enhancement and the Initial Mass Function of
%Extremely Metal-poor Stars in the Galactic Halo
%\bibitem[Komiya et al.(2007)]{Komiya:07} Komiya, Y., Suda, T., 
%Minaguchi, H., et al.\ 2007, \apj, 658, 367 

%DONTCITE
%Early-Age Evolution of the Milky Way Related by Extremely Metal-Poor
%Stars
%\bibitem[Komiya et al.(2009)]{Komiya:09} Komiya, Y., Suda, T., 
%\& Fujimoto, M.~Y.\ 2009, \apj, 694, 1577 


%The [Fe/H], [C/Fe], and [α/Fe] Distributions of the Boötes I Dwarf
%Spheroidal Galaxy
\bibitem[Lai et al.(2011)]{Lai:11} Lai, D.~K., Lee, Y.~S., 
Bolte, M., et al.\ 2011, \apj, 738, 51 

%CHEMOBS
%A Mass-dependent Yield Origin of Neutron-capture Element Abundance
%Distributions in Ultra-faint Dwarfs
\bibitem[Lee et al.(2013)]{Lee:13} Lee, D.~M., Johnston, 
K.~V., Tumlinson, J., Sen, B., \& Simon, J.~D.\ 2013, \apj, 774, 103 

%Molecular processes and gravitational collapse in intergalactic shocks
\bibitem[Mac Low 
\& Shull(1986)]{MacLow:86} Mac Low, M.-M., \& Shull, J.~M.\ 1986, \apj, 302, 585 



%Rotating massive stars: From first stars to gamma ray bursts
\bibitem[Maeder 
\& Meynet(2012)]{Meader:12} Maeder, A., \& Meynet, G.\ 2012, Reviews of Modern Physics, 84, 25 

%Revisiting the First Galaxies: The Epoch of Population III Stars
\bibitem[Muratov et al.(2013)]{Muratov:13} Muratov, A.~L., Gnedin, 
O.~Y., Gnedin, N.~Y., \& Zemp, M.\ 2013, \apj, 773, 19 




%DONTCITE
%Metal Enrichment of the Primordial Interstellar Medium through
%Three-dimensional Hydrodynamical Evolution of the First Supernova
%Remnant
%\bibitem[Nakasato 
%\& Shigeyama(2000)]{Nakasato:00} Nakasato, N., \& Shigeyama, T.\ 2000, \apjl, 541, L59

%DONTCITE
%Metal enrichment history of the proto-galactic interstellar medium
%\bibitem[Nakasato(2002)]{Nakasato:02} Nakasato, N.\ 2002, ApSS, 281, 257 

%Nucleosynthesis in Stars and the Chemical Enrichment of Galaxies
\bibitem[Nomoto et 
al.(2013)]{Nomoto:13} Nomoto, K., Kobayashi, C., \& Tominaga, N.\ 2013, \araa, 51, 457

%An Extremely Carbon-rich, Extremely Metal-poor Star in the Segue 1 System
\bibitem[Norris et al.(2010)]{Norris:10a} Norris, J.~E., Gilmore, 
G., Wyse, R.~F.~G., Yong, D., \& Frebel, A.\ 2010, \apjl, 722, L104 

%DONTCITE
%Chemical Enrichment in the Faintest Galaxies: The Carbon and Iron
%Abundance Spreads in the Boötes I Dwarf Spheroidal Galaxy and the
%Segue 1 System
%\bibitem[Norris et al.(2010)]{Norris:10b} Norris, J.~E., Wyse, 
%R.~F.~G., Gilmore, G., et al.\ 2010, \apj, 723, 1632 


%Boo-1137—an Extremely Metal-Poor Star in the Ultra-Faint Dwarf
%Spheroidal Galaxy Boötes I
\bibitem[Norris et al.(2010)]{Norris:10c} Norris, J.~E., Yong, D., 
Gilmore, G., \& Wyse, R.~F.~G.\ 2010, \apj, 711, 350 

%A New Look at Simple Inhomogeneous Chemical Evolution
\bibitem[Oey(2000)]{Oey:00} Oey, M.~S.\ 2000, \apjl, 542, L25

%The number and metallicities of the most metal-poor stars
\bibitem[Oey(2003)]{Oey:03} Oey, M.~S.\ 2003, \mnras,
  339, 849 

\bibitem[Oh \& Haiman(2002)]{Oh:02} Oh, S.~P., \& Haiman, Z.\ 2002, \apj, 569, 558 

%The Ultraviolet Luminosity Function of the Earliest Galaxies
\bibitem[O'Shea et al.(2015)]{OShea:15} O'Shea, B.~W., Wise, 
J.~H., Xu, H., \& Norman, M.~L.\ 2015, arXiv:1503.01110 



%Mixing in Supersonic Turbulence
\bibitem[Pan 
\& Scannapieco(2010)]{Pan:10} Pan, L., \& Scannapieco, E.\ 2010, \apj, 721, 1765

%DONTCITE
%Modeling the Pollution of Pristine Gas in the Early Universe
%\bibitem[Pan et al.(2013)]{Pan:13} Pan, L., Scannapieco, E., 
%\& Scalo, J.\ 2013, \apj, 775, 111 

%DONTCITE
%Near-pristine gas at high redshifts: a window on early nucleosynthesis
%\bibitem[Pettini 
%\& Cooke(2012)]{2012nuco.confE..71P} Pettini, M., \& Cooke, R.\ 2012,
%Nuclei in the Cosmos (NIC XII), Cairns, Australia

%Determination of Nucleosynthetic Yields of Supernovae and Very
%Massive Stars from Abundances in Metal-Poor Stars
\bibitem[Qian 
\& Wasserburg(2002)]{Qian:02} Qian, Y.-Z., \& Wasserburg, G.~J.\ 2002, \apj, 567, 515 

%Abundances of Sr, Y, and Zr in Metal-Poor Stars and Implications for Chemical Evolution in the Early Galaxy
\bibitem[Qian 
\& Wasserburg(2008)]{Qian:08} Qian, Y.-Z., \& Wasserburg, G.~J.\ 2008, \apj, 687, 272 

%DONTCITE
%Supernova-driven outflows and chemical evolution of dwarf spheroidal galaxies
%\bibitem[Qian 
%\& Wasserburg(2012)]{Qian:12} Qian, Y.-Z., \& Wasserburg, G.~J.\ 2012, Proceedings of the National Academy of Science, 109, 4750 

%The Fate of the First Galaxies. III. Properties of Primordial Dwarf
%Galaxies and Their Impact on the Intergalactic Medium
\bibitem[Ricotti et al.(2008)]{Ricotti:08} Ricotti, M., Gnedin, 
N.~Y., \& Shull, J.~M.\ 2008, \apj, 685, 21 

%Confined Population III Enrichment and the Prospects for Prompt
%Second-generation Star Formation
\bibitem[Ritter et al.(2012)]{Ritter:12} Ritter, J.~S., 
Safranek-Shrader, C., Gnat, O., Milosavljevi{\'c}, M., 
\& Bromm, V.\ 2012, \apj, 761, 56 

%DONTCITE
%Neutron-capture Nucleosynthesis in the First Stars
%\bibitem[Roederer et al.(2014)]{Roederer:14} Roederer, I.~U., 
%Preston, G.~W., Thompson, I.~B., Shectman, S.~A., 
%\& Sneden, C.\ 2014, \apj, 784, 158 

%DONTCITE
%Chemical enrichment in very low metallicity environments: Boötes I
%\bibitem[Romano et al.(2015)]{Romano:15} Romano, D., Bellazzini, 
%M., Starkenburg, E., \& Leaman, R.\ 2015, \mnras, 446, 4220 



%Star formation in the first galaxies - I. Collapse delayed by
%Lyman-Werner radiation
\bibitem[Safranek-Shrader et al.(2012)]{SafranekShrader:12} 
Safranek-Shrader, C., Agarwal, M., Federrath, C., et al.\ 2012, \mnras, 
426, 1159 

%Star formation in the first galaxies - II. Clustered star formation and the influence of metal line cooling
\bibitem[Safranek-Shrader et al.(2014)]{SafranekShrader:14a} 
Safranek-Shrader, C., Milosavljevi{\'c}, M., 
\& Bromm, V.\ 2014, \mnras, 438, 1669 


%DONTCITE
%Formation of the first low-mass stars from cosmological initial conditions
%\bibitem[Safranek-Shrader et al.(2014)]{SafranekShrader:14} 
%Safranek-Shrader, C., Milosavljevi{\'c}, M., 
%\& Bromm, V.\ 2014, \mnras, 440, L76 

%DONTCITE
%\bibitem[Safranek-Shrader et al.(2015)]{2015arXiv150103212S} 
%Safranek-Shrader, C., Montgomery, M., Milosavljevic, M., 
%\& Bromm, V.\ 2015, arXiv:1501.03212 


%DONTCITE
%Early star formation, nucleosynthesis and chemical evolution in proto-galactic clouds
%\bibitem[Saleh et al.(2006)]{Saleh:06} Saleh, L., Beers, T.~C., 
%\& Mathews, G.~J.\ 2006, Journal of Physics G Nuclear Physics, 32, 681 

%Ultra faint dwarfs: probing early cosmic star formation
\bibitem[Salvadori 
\& Ferrara(2009)]{Salvadori:09} Salvadori, S., \& Ferrara, A.\ 2009, \mnras, 395, L6 

%DONTCITE
%First stars in damped Lyα systems
%\bibitem[Salvadori 
%\& Ferrara(2012)]{Salvadori:12} Salvadori, S., \& Ferrara, A.\ 2012, \mnras, 421, L29 

%Hydrogen molecules and the radiative cooling of pregalactic shocks
\bibitem[Shapiro 
\& Kang(1987)]{Shapiro:87} Shapiro, P.~R., \& Kang, H.\ 1987, \apj, 318, 32 



%The History of R-Process Enrichment in the Milky Way
\bibitem[Shen et al.(2014)]{Shen:14} Shen, S., Cooke, R., 
Ramirez-Ruiz, E., et al.\ 2014, arXiv:1407.3796 

%Fossil Imprints of the First-Generation Supernova Ejecta in Extremely Metal-deficient Stars
\bibitem[Shigeyama 
\& Tsujimoto(1998)]{Shigeyama:98} Shigeyama, T., \& Tsujimoto,
T.\ 1998, \apjl, 507, L135

% DONTCITE
% %Excavation of the First Stars
% \bibitem[Shigeyama et al.(2003)]{2003} Shigeyama, T., 
% Tsujimoto, T., \& Yoshii, Y.\ 2003, \apjl, 586, L57 

% DONTCITE
%High-resolution Spectroscopy of Extremely Metal-poor Stars in the
%Least Evolved Galaxies: Leo IV
%\bibitem[Simon et al.(2010)]{Simon:10} Simon, J.~D., Frebel, A., 
%McWilliam, A., Kirby, E.~N., \& Thompson, I.~B.\ 2010, \apj, 716, 446 

%Chemical Signatures of the First Supernovae in the Sculptor Dwarf Spheroidal Galaxy
\bibitem[Simon et al.(2015)]{Simon:15} Simon, J.~D., Jacobson, 
H.~R., Frebel, A., et al.\ 2015, \apj, 802, 93 



%Preparing for an Explosion: Hydrodynamic Instabilities and Turbulence in Presupernovae
\bibitem[Smith 
\& Arnett(2014)]{Smith:14} Smith, N., \& Arnett, W.~D.\ 2014, \apj, 785, 82 

%Neutron-Capture Elements in the Early Galaxy
\bibitem[Sneden et 
al.(2008)]{Sneden:08} Sneden, C., Cowan, J.~J., \& Gallino, R.\ 2008, \araa, 46, 241 

% DONTCITE
%Carbon-enhanced metal-poor stars: the most pristine objects?
%\bibitem[Spite et 
%al.(2013)]{Spite:14} Spite, M., Caffau, E., Bonifacio, P., et al.\ 2013, \aap, 552, A107 

%The first stars: formation of binaries and small multiple systems
\bibitem[Stacy et al.(2010)]{Stacy:10} Stacy, A., Greif, T.~H., 
\& Bromm, V.\ 2010, \mnras, 403, 45 

%The first stars: mass growth under protostellar feedback
\bibitem[Stacy et al.(2012)]{Stacy:12} Stacy, A., Greif, T.~H., 
\& Bromm, V.\ 2012, \mnras, 422, 290 

% DONTCITE
%The Stellar Abundances for Galactic Archaeology (SAGA) data base -
%II. Implications for mixing and nucleosynthesis in extremely
%metal-poor stars and chemical enrichment of the Galaxy
%\bibitem[Suda et al.(2011)]{Suda:11} Suda, T., Yamada, S., 
%Katsuta, Y., et al.\ 2011, \mnras, 412, 843 

%The Evolution of Galaxies. II. Chemical Evolution Coefficients
\bibitem[Talbot 
\& Arnett(1973)]{Talbot:73} Talbot, R.~J., Jr., \& Arnett, W.~D.\ 1973, \apj, 186, 51 


%Ultra-faint Dwarf Galaxies as a Test of Early Enrichment and
%Metallicity-dependent Star Formation
\bibitem[Tassis et al.(2012)]{Tassis:12} Tassis, K., Gnedin, N.~Y., \& Kravtsov, A.~V.\ 2012, \apj, 745, 68 

%Principal component analysis on chemical abundances spaces
\bibitem[Ting et al.(2012)]{Ting:12} Ting, Y.-S., Freeman, 
K.~C., Kobayashi, C., De Silva, G.~M., 
\& Bland-Hawthorn, J.\ 2012, \mnras, 421, 1231

% DONTCITE
%Abundance Profiling of Extremely Metal-poor Stars and Supernova
%Properties in the Early Universe
%\bibitem[Tominaga et al.(2014)]{Tominaga:14} Tominaga, N., Iwamoto, 
%N., \& Nomoto, K.\ 2014, \apj, 785, 98

% DONTCITE
% %A new interpretation of the heavy element abundances in metal-deficient stars
% \bibitem[Truran(1981)]{Truran:81} Truran, J.~W.\ 1981, \aap, 97, 391

%Probing the Neutron-Capture Nucleosynthesis History of Galactic
%Matter
\bibitem[Truran et al.(2002)]{Truran:02} Truran, J.~W., Cowan, 
J.~J., Pilachowski, C.~A., \& Sneden, C.\ 2002, \pasp, 114, 1293 

%New Insights into the Early Stage of the Galactic Chemical Evolution
\bibitem[Tsujimoto 
\& Shigeyama(1998)]{Tsujimoto:98} Tsujimoto, T., \& Shigeyama, T.\ 1998, \apjl, 508, L151 

%Chemical Evolution of the Galactic Halo through Supernova-induced
%Star Formation and Its Implication for Population III Stars
\bibitem[Tsujimoto et al.(1999)]{Tsujimoto:99} Tsujimoto, T., 
Shigeyama, T., \& Yoshii, Y.\ 1999, \apjl, 519, L63 

\bibitem[Turk et al.(2011)]{Turk:11} Turk, M.~J., Smith, B.~D., 
Oishi, J.~S., et al.\ 2011, \apjs, 192, 9 

%DONTCITE
% %Relics of Metal-free Low-Mass Stars Exploding as Thermonuclear
% %Supernovae
% \bibitem[Tsujimoto 
% \& Shigeyama(2006)]{Tsujimoto:06} Tsujimoto, T., \& Shigeyama, T.\ 2006, \apjl, 638, L109 

%DONTCITE
% %Chemical Evolution in Hierarchical Models of Cosmic Structure. I. Constraints on the Early Stellar Initial Mass Function
% \bibitem[Tumlinson(2006)]{Tumlinson:06} Tumlinson, J.\ 2006, \apj, 
% 641, 1 
	
%DONTCITE
%Chemical Evolution in Hierarchical Models of Cosmic Structure. II. The Formation of the Milky Way Stellar Halo and the Distribution of the Oldest Stars
%\bibitem[Tumlinson(2010)]{Tumlinson:10} Tumlinson, J.\ 2010, \apj, 
%708, 1398 

% The Distribution of Alpha Elements in Ultra-faint Dwarf Galaxies
\bibitem[Vargas et al.(2013)]{Vargas:13} Vargas, L.~C., Geha, M., 
Kirby, E.~N., \& Simon, J.~D.\ 2013, \apj, 767, 134 

%DONTCITE
% %Stellar Chemical Signatures and Hierarchical Galaxy Formation
% \bibitem[Venn et al.(2004)]{Venn:04} Venn, K.~A., Irwin, M., 
% Shetrone, M.~D., et al.\ 2004, \aj, 128, 1177 

%DONTCITE
% CHEMMOD
%Chemical evolution of classical and ultra-faint dwarf spheroidal galaxies
%\bibitem[Vincenzo et al.(2014)]{Vincenzo:14} Vincenzo, F., 
%Matteucci, F., Vattakunnel, S., 
%\& Lanfranchi, G.~A.\ 2014, MNRAS, 441, 2815 

%Galactic r-process enrichment by neutron star mergers in cosmological
%simulations of a Milky Way-mass galaxy
\bibitem[van de Voort et al.(2014)]{Voort:14} van de Voort, F., 
Quataert, E., Hopkins, P.~F., Keres, D., 
\& Faucher-Giguere, C.-A.\ 2014, arXiv:1407.7039 

%Ultrafaint Dwarfs?Star Formation and Chemical Evolution in the Smallest Galaxies
\bibitem[Webster et al.(2014)]{Webster:14} Webster, D., 
Sutherland, R., \& Bland-Hawthorn, J.\ 2014, \apj, 796, 11 

%DONTCITE
%Star Formation in Ultrafaint Dwarfs: Continuous or Single-Age Bursts?
%\bibitem[Webster et al.(2015)]{Webster:15} Webster, D., 
%Bland-Hawthorn, J., \& Sutherland, R.\ 2015, \apjl, 799, LL21 


%DONTCITE
% %Abundance ratios as a function of metallicity
% \bibitem[Wheeler et 
%  al.(1989)]{Wheeler:89} Wheeler, J.~C., Sneden, C., \& Truran, J.~W., Jr.\ 1989, \araa, 27, 279 

%DONTCITE
%Stochastic effects in the chemical evolution of galaxies
%\bibitem[White 
%\& Audouze(1983)]{White:83} White, S.~D.~M., \& Audouze, J.\ 1983, \mnras, 203, 603 

%Resolving the Formation of Protogalaxies. III. Feedback from the
%First Stars
\bibitem[Wise 
\& Abel(2008)]{Wise:08} Wise, J.~H., \& Abel, T.\ 2008, \apj, 685, 40 

%The Birth of a Galaxy: Primordial Metal Enrichment and Stellar
%Populations
\bibitem[Wise et al.(2012)]{Wise:12} Wise, J.~H., Turk, M.~J., 
Norman, M.~L., \& Abel, T.\ 2012, \apj, 745, 50 

%The birth of a galaxy ? III. Propelling reionization with the faintest galaxies
\bibitem[Wise et al.(2014)]{Wise:14} Wise, J.~H., Demchenko, 
V.~G., Halicek, M.~T., et al.\ 2014, \mnras, 442, 2560 



%Three-dimensional neutrino-driven supernovae: Neutron star kicks,
%spins, and asymmetric ejection of nucleosynthesis products
\bibitem[Wongwathanarat et 
al.(2013)]{Wongwathanarat:13} Wongwathanarat, A., Janka, H.-T., M\"uller, E.\ 2013, \aap, 552, A126 

%The Most Metal-poor Stars. II. Chemical Abundances of 190 Metal-poor Stars Including 10 New Stars with [Fe/H] <= -3.5
\bibitem[Yong et al.(2013)]{Yong:13a} Yong, D., Norris, J.~E., 
Bessell, M.~S., et al.\ 2013, \apj, 762, 26 


%DONTCITE
% %The Most Metal-poor Stars. III. The Metallicity Distribution Function
% %and Carbon-enhanced Metal-poor Fraction
% \bibitem[Yong et al.(2013)]{Yong:13} Yong, D., Norris, J.~E., 
% Bessell, M.~S., et al.\ 2013, \apj, 762, 27 

\bibitem[Zucker et al.(2012)]{Zucker:12} Zucker, D.~B., de Silva, 
G., Freeman, K., Bland-Hawthorn, J., 
\& Hermes Team 2012, Galactic Archaeology: Near-Field Cosmology and the Formation of the Milky Way, 458, 421 




\end{thebibliography}
\end{document}